\documentclass[12pt,preprint]{aastex}

\shorttitle{Instabilities in Neutron Stars}

\shortauthors{Ou \& Tohline }

\begin{document}


\title{{Unexpected Dynamical Instabilities In Differentially Rotating Neutron Stars}}


\author{Shangli Ou and Joel E. Tohline}
\affil{
Department of Physics \& Astronomy,
Center for Computation \& Technology, \\
 Louisiana State University, Baton Rouge, LA  70803}

%\affil{Department of Physics \& Astronomy, Louisiana State
%University, Baton Rouge, LA  70803}

%\and
%
%\author{Lee Lindblom}
%\affil{Theoretical Astrophysics 130-33,
%       California Institute of Technology, Pasadena, CA 91125}


\begin{abstract}
A one-armed spiral instability has been found to develop in
differentially rotating stellar models that have a relatively stiff,
$n=1$ polytropic equation of state and a wide range of rotational
energies. This suggests that such instabilities can arise in neutron
stars that are differentially, although not necessarily rapidly,
rotating. The instability seems to be directly triggered by the
presence of a corotation resonance inside the star. Our analysis
also suggests that a resonant cavity resulting from a local minimum
in the radial vortensity profile of the star plays an important role
in amplifying the unstable mode.  Hence, it appears as through this
instability is closely related to the so-called ``Rossby wave
instability'' \citep{LLCN99} that has been found to arise in
accretion disks. In addition to the one-armed ($m=1$) spiral mode,
we have found that higher-order ($m = 2$ and $m=3$) nonaxisymmetric
modes also can become unstable if corotation points that resonate
with the eigenfrequencies of these higher-order modes also appear
inside the star. The growth rate of each mode seems to depend on the
location of its corotation radius with respect to the vortensity
profile (or on the depth of its corotation radius inside the
vortensity well). The existence of such instabilities makes the
stability criterion for differentially rotating neutron stars
non-unique. Also, the gravitational-waves emitted from such unstable
systems generally will not have a monochromatic frequency spectrum.

\end{abstract}


%% subject header.  Objects should be in the appropriate "individual"
%% headers (e.g. quasars: individual, stars: individual, etc.) with the
%% additional provision that the total number of headers, including each
%% individual object, not exceed six.  The \objectname{} macro, and its
%% alias \object{}, is used to mark each object.  The macro takes the object
%% name as its primary argument.  This name will appear in the paper
%% and serve as the link's anchor in the electronic edition if the name
%% is recognized by the data centers.  The macro also takes an optional
%% argument in parentheses in cases where the data center identification
%% differs from what is to be printed in the paper.



\section{Introduction}

Rotating stars are subject to a variety of nonaxisymmetric
instabilities \citep{Ch69, T78, A03}.
If a star rotates sufficiently fast -- to a point
where the ratio of its rotational to gravitational potential
energy $T/|W| \gtrsim 0.27$ -- it will encounter a dynamical
bar-mode instability that will result in the deformation of the
star into a rapidly spinning, bar-like structure. This instability
has been verified by a number of numerical hydrodynamics
investigations \citep{TDM85, DGTB86, WT88, CT00, NCT00, B00,L02}. At a
slower rotation rate, $T/|W| \gtrsim 0.14$, a star may also
encounter a secular bar-mode instability that can promote
deformation into a bar-like shape, but only if the star is
subjected to a dissipative process capable of redistributing
angular momentum within its structure, such as viscosity or
gravitational radiation reaction (GRR) forces
\citep{Ch70,FS78,IL90,IL91,LS95}. In reality, viscosity and GRR
forces tend to compete with each other and can drive the secularly
unstable star along a variety of different evolutionary paths.
%%Recently,
%%\cite{OTL04} have used numerical hydrodynamic technique to study
%%the non-linear evolution of the GRR-driven bar-mode instability
%%in rigidly rotating $\mathrm{n}=1/2$ neutron stars in the
%%absence of any competing dissipation, where
%%$\mathrm{n}$ is the polytropic index of the simulated star.
%%At the same time, \cite{SK04} have also conducted a similar
%%study on the secular bar-mode instability induced by GRR
%%in differentially rotating $\mathrm{n}=1$ neutron stars.
Recently, \cite{SK04} and \cite{OTL04} used numerical hydrodynamic
techniques to study the nonlinear evolution of the GRR-driven
secular bar-mode instability in rotating, ``polytropic'' neutron
stars in the absence of any competing dissipation. The above
studies have shown that the critical limits of both the secular
and dynamical bar-mode instabilities do not depend sensitively on
the degree of compressibility of the equation of state (EOS) or on
the differential rotation law of a star.

Surprisingly, several recent numerical studies have found that
dynamical instabilities that excite $m=1$ as well as $m=2$ azimuthal
Fourier modes can arise in self-gravitating structures having $T/|W|
< 0.27$, that is, having a rotational energy below the threshold
traditionally expected for dynamical nonaxisymmetirc instabilities.
\cite{TH90} found an $m=2$ instability in self-gravitating rings and
tori having uniform specific angular momentum and $T/|W| \gtrsim
0.16$. \cite{CNLB01} found a one-armed ($m=1$) spiral instability in
stellar models with a polytropic index $n=3.33$, strong differential
rotation and $T/|W| \sim 0.14$. \cite{SKE02, SKE03} found an $m=2$
instability in extremely differentially rotating stellar models
having $T/|W|$ on the order of $10^{-2}$! \cite{saijo:03} concluded
that one-armed ($m=1$) modes only appear in structures with a very
soft equation of state ($n \gtrsim 2.5$), and that a necessary
condition for this instability to arise is the presence of an
off-center density maximum, that is, a toroidal-like structure.
\cite{ott:05a} showed that a one-armed spiral instability may arise in
the context of a pre-supernova core collapse. The linear analysis of
\cite{watts:04} has suggested that these low $T/|W|$ instabilities
are triggered because corotation points of the unstable modes
fall within the differentially rotating structure of the examined
models. \cite{saijo:05} also have carried out a linear stability
analysis that shows a connection between the one-armed, $m=1$
instability and corotation points in differentially rotating stars.

Many earlier studies of disks and tori have also noted an
association between the onset of dynamical instabilities and the
existence of corotation points. For example, a number of groups
\citep{PP84,GGN86,NGG87,FR88} have used both linear analyses and
nonlinear simulations to study the so-called Papaloizou-Pringle
instability, which is triggered by corotation resonances. But the
existence of a corotation point within a model is not sufficient
to ensure significant growth of an unstable mode. It appears as
though a resonant cavity is also required to drive these modes to
very large amplitude. In the case of the Papaloizou-Pringle
instability, the inner and outer edges of the disk or torus form a
resonant cavity in which waves are reflected back and forth; the
unstable mode is greatly amplified through multiple passages
across the corotation radius. However, it is not obvious how such
a resonant cavity can exist in stellar models that have no inner
edges.

\cite{LH78} and \cite{PL89}
have discussed how the radial distribution of vortensity
or potential vorticity, which is defined as the ratio
between vorticity and surface density, is
important in determining the stability of homentropic,
self-gravitating rings and disks. \cite{LLCN99} and \cite{LFLC00}
extended this discussion to non-homentropic disks. According to
their analysis of the so-called ``Rossby-Wave Instability'',
local minima in the radial vortensity profile
of homentropic configurations 
tend to trap radially propagating waves and, hence, can serve as
resonant cavities that will help drive corotation instabilities to
large amplitude. This removes the necessity of having disk edges
to serve as reflecting boundaries. In their subsequent nonlinear
simulations, \cite{LCWL01} found that this mechanism can aid in
the excitation of nonaxisymmetric instabilities ($m=1$,
$3$, $5$) and vortex formation in accretion disks. It is
interesting to note that when the initial perturbations in these
nonlinear simulations contained the pure eigenfunctions for a
given azimuthal mode (for example, $m=3$ or $5$), the
corresponding mode developed; but, when a white noise initial
perturbation was used, an $m=1$ spiral mode became
dominant in the final stage.

It is conceivable that a similar resonant cavity mechanism may work
to amplify corotation resonances in stellar models. This would seem
to be especially likely for rapidly rotating models that have
toroidal-like structures because the associated off-center density
maximum will inevitably result in a dip in the radial vortensity
profile of the star.  This profile could then act as a resonant
cavity to trap radially propagating waves. Alternatively, a maximum
in the star's radial entropy profile can also produce such a cavity
\citep{LLCN99}, but in keeping with most previous studies of the low
$T/|W|$ instabilities we will confine our analysis to homentropic
configurations where a dip in the vortensity profile will be
required to create a resonant cavity.

In this paper, we report the unexpected discovery of an
$m=1$ spiral-mode instability in rotating stars with a
relatively stiff ($n=1$) polytropic EOS that was designed
to represent differentially rotating neutron stars. We also show
evidence that this instability is directly triggered by
the existence inside the star of a corotation radius for the
eigenmode. This one-armed spiral instability develops on a
timescale that is somewhat longer than the growth time observed in
previous studies of configurations with a softer EOS, but the
instability appears to be dynamical nevertheless.
%%and are sufficient to surpass the secular bar-mode instability.
%%The final results are not sensitive to the rotation law.
All of our unstable models exhibit a local minimum in their radial
vortensity profile and this seems likely to be the resonant cavity
that traps the wave and permits it to amplify. (We have found that
such a minimum can exist in configurations with centrally
condensed, rather than toroidal, density structures when the
degree of differential rotation is sufficiently strong.) Our
results suggest that it is the combination of corotation points
and resonant cavities formed by vortensity minima that is
responsible for the excitation of low $T/|W|$ instabilities found
in stellar models.

Perhaps our most striking discovery is that, in addition to the
$m=1$ mode, $m$=2 and 3 modes also come into
play as long as their corotation points exist within the structure
of the star. Therefore, if the rotational profile of the star is
steep enough to contain corotation points of multiple modes, these
unstable modes may all arise within one single star. The entire
picture of rotational instabilities that might arise in
differentially rotating neutron stars becomes more complicated due
to the co-existence of these corotation instabilities. The
corresponding gravitational-wave signals may exhibit
different characteristic frequencies at different or overlapping times,
since various unstable modes may set in simultaneously.

%%The GWs from such a scenario have unique characteristics and may
%%be observable by GW detectors, such as LIGO.
\begin{deluxetable}{ccccccccccc}
\tablewidth{0pt}
\tablecaption{Parameters of Initial Models }
\tablehead{ \colhead{Model} & \colhead{$n$} & \colhead{$A/R_{eq}$} & \colhead{$R_p/R_{eq}$} &
  \colhead{$M_{tot}$} & \colhead{$\rho_c/\rho_{max}$} & \colhead{$\bar{\rho}$} & \colhead{$\omega_c$} &
   \colhead{$T/|W|$} & \colhead{unstable modes} }
\startdata
%%   V254 & 1.0 & -- & 0.309 & 0.1879& 0.9831 & 0.3709& 4.28  & 0.254 & m=1\\
   J250 & 1.0 & 0.94 & 0.302 & 0.204 & 0.976 & 0.384 & 1.44 & 0.250 & m=1\\
   J133 & 1.0 & 0.94 & 0.558 & 0.257 & 0.999 &0.309 &  1.19 & 0.133 & - \\
   J127 & 1.0 & 0.59 & 0.558 &0.335 & 0.999 & 0.363& 1.75 & 0.127 & m=1,2,3\\
   J068 & 1.0 & 0.44 & 0.721 & 0.384&  1.000 & 0.343& 1.80  & 0.068 & m=1,2\\
\enddata
\label{table:model} 
\tablecomments{
   $n$ is polytropic index, A is a parameter that determines the degree of 
   differential rotation, $R_{eq}$ is the equatorial radius, $R_p$ is the polar radius,
   $M_{tot}$ is the total mass, $\rho_c$ is the central density, $\rho_{max}$ is the 
   maximum density, $\bar{\rho}$ is the mean density, $\omega_c$ is the angular frequency
   at the center, $T$ is the rotational energy, $W$ is the gravitational energy.}
\end{deluxetable}

\section{Initial Models and Numerical Methods}

\subsection{Initial Axisymmetric Models}

Using the Hachisu self-consistent-field technique \citep{H86}, four
simplified axisymmetric neutron star models (models {\bf J250}, {\bf
J133}, {\bf J127}, and {\bf J068} ) were constructed on a
cylindrical grid with resolution $66 \times 82 \times 128$ in the
radial, vertical and azimuthal directions, respectively. For each
model, we have adopted a dimensionless unit system in which the
gravitational constant $G$, the maximum density
$\rho_{\mathrm{max}}$, and the radius of the entire grid
$R_{\mathrm{grid}}$ are set to 1. For comparison purposes,
equatorial radii  $R_{\mathrm{eq}}$ were all set to 0.673. All the
models were constructed using Newtonian gravity and a polytropic
equation of state, $p=K\rho^{1+1/n}$, where $p$ is the gas pressure,
$K$ is a polytropic constant, and the polytropic index $n$ was in
all cases chosen to be 1. The angular velocity profiles were
specified by the so-called j-constant-like law as,
\begin{equation}
   \Omega(R) = \frac{\omega_c A^2}{R^2 + A^2} \, , \label{EQ:jconst}
\end{equation}
\noindent where $R= \sqrt{x^2+y^2}$ is the cylindrical
radius, $A$ is a constant, and $\omega_c$ is the angular velocity
at the rotation axis (the Z-axis).
%%Model ``V254" belongs to the so-called ``uniform vortensity" models
%%\cite{CT00} have used to study the evolution of protostellar gas
%%clouds.
The specific parameter values of each model are given in Table
\ref{table:model}, where $\bar{\rho}$ is the mean density of the
star, $\rho_c$ is the central density (which will be less than
$\rho_{\mathrm{max}}$ if the model exhibits an off-center density
maximum), and $R_p$ is the polar radius.

\subsection{Analysis Method}
We used a three-dimensional (3D), Newtonian hydrodynamics code
essentially the same as the
one employed by \citet[hereafter Paper I]{OTL04} to study the
evolution of our differentially rotating, neutron stars models.
The details of the hydrodynamics code can be found in \cite{MTF02}.
In paper I, an artificially enhanced gravitational-radiation-reaction
(GRR) potential
was introduced into the code to mimic the GRR effect,
which will induce the secular bar-mode instability
when the $T/|W|$ value of the star exceeds the critical limit 0.14.

In order to
monitor the growth of nonaxisymmetric modes, we measured the
Fourier amplitude of each mode with azimuthal quantum number
$m$ in the following fashion: At any point in time during an
evolution, the azimuthal density
distribution in a ring of fixed $R$ and $z$
can be decomposed into a series of azimuthal Fourier components
via the relation,
\begin{equation}
  \rho(R,z, \phi) = \sum\limits_{m=-\infty}^{+\infty}
      C_m(R,z) e^{i m \phi} \, ,
\end{equation}
where the complex amplitudes $C_m$ are defined by the expression,
\begin{equation}
   C_m(R,z) = \frac{1}{2\pi} \int_0^{2\pi}
      \rho(R,z, \phi) e^{-i m \phi} d \phi \,.
\end{equation}
In our simulations, the time-dependent behavior of the magnitude of
this coefficient, $|C_m|$, is then monitored to measure the growth-rate
of various unstable modes. At any point in time, the ratio between 
the imaginary and real parts of $C_m(R,z)$ gives us a spatially
dependent phase angle $\phi_m(R, z)$ from which we also are able to
determine the azimuthal structure of each unstable mode $m$,
if it exists.
(In particular, the
radial dependence of $\phi_m$ tells us, for example, whether an
$m=1$ mode has a spiral character or an $m=2$ mode has a bar-like
structure; see Figures \ref{fig:phase_j250} and
\ref{fig:phase_j068}, below.)
In general, $\phi_m(R, z)$ is a function of time,  
but an eigenmode of the system becomes
identifiable when the spatial structure of the phase angle
$\phi_m(R, z)$ appears to be coherent/aligned throughout the star
and time-independent, apart from a spatially independent azimuthal
eigenfrequency $\omega_m \equiv \partial \phi_m/\partial t$ of the mode. 
Another way to look at this is that the oscillation period of mode $m$,
$T_m\equiv 2\pi/\omega_m$, is equivalent to the time it takes the
real or imaginary parts of $C_m$ to complete a full cycle from a
positive value to a negative value, then back to a positive value.
In our nonlinear hydrodynamic simulations, the measurement of
$\omega_m$ is accurate only when the corresponding mode dominates
the evolution.

By definition, the corotation radius $R_\mathrm{cor}$ of mode
$m$ is the radial location in the star where the angular
velocity of the fluid resonates with the eigenfrequency of the
unstable mode, that is, where $\omega (R_{cor}) =\omega_m/m$.
In our analysis, we also will find it useful to refer to each
model's central spin period, $T_c=2 \pi /\omega_c$, and
characteristic dynamical time scale, $T_0\equiv 2 \pi/\omega_0$,
where $\omega_0 \equiv \sqrt{\pi G\bar{\rho}}$.
\section{Simulation Results}

\begin{deluxetable}{cccccccccc}
\tablewidth{0pt}
\tablecaption{Relevant Frequencies of All Models \label{table:freq}}
\tablehead{ \colhead{Model} & \colhead{$\omega_c$} & \colhead{$\omega_s$} &
   \colhead{$\omega_0= \sqrt{ \pi G \bar{\rho}}$} &
  \colhead{$\omega_1$} & \colhead{$\omega_2/2$} & \colhead{$\omega_3/3$} &
   \colhead{$\tau_1 /T_0$} & \colhead{$\tau_2 /T_0$} &
   \colhead{$\tau_3 /T_0$} }
\startdata
%%   V254 & 4.28 & 0.38 & 1.08 & 1.09 &   -  &  -  & 11.3 &  -   &-\\
%%   J250 & 1.44 & 0.66 & 1.10 & 1.15 &   -  &  -  & 9.4  &  -   &- \\
%%   J133 & 1.19 & 0.54 & 0.98 &   -  &   -  &  -  &  -   &  -   &- \\
%%   J127 & 1.75 & 0.44 & 1.07 & 1.22 & 0.89 & 0.98& 29.4 & 7.84 & 4.2 \\
%%   J068 & 1.80 & 0.29 & 1.04 & 1.34 &  0.9 &  -  &  8.4& 4.39 &-    \\
   J250 & 1.44 & 0.66 & 1.10 & 1.2 &   -  &  -  & 9.4  &  -   &- \\
   J133 & 1.19 & 0.54 & 0.98 &   -  &   -  &  -  &  -   &  -   &- \\
   J127 & 1.75 & 0.44 & 1.07 & 1.3 & 0.9 & 1.0& 29.4 & 7.8 & 4.2 \\
   J068 & 1.80 & 0.29 & 1.04 & 1.3 &  0.9 &  -  &  8.4& 4.4 &-    \\
\enddata
\tablecomments{
   $\omega_c$ is the angular frequency at the center, 
   $\omega_s$ is the angular frequency at the equatorial surface, 
   $\bar{\rho}$ is the mean density, 
   $\omega_m$ is the eigenfrequency of the azimuthal mode $m$,
   $\tau_m$ is the e-folding time of the azimuthal mode $m$, $T_0= 2\pi/\omega_0$.
   As illustrated in Figure \ref{fig:freq_j127}, for a given model,
   the eigenfrequencies $\omega_m$ have been measured at different evolutionary times,
   when each mode dominates.
}
\end{deluxetable}

In this section, we show results from a number of independent 3D
hydrodynamic evolutions. Because our initial goal was to study the
secular bar-mode instability induced by GRR forces, rather than the
development of a one-armed spiral instability, the initially
axisymmetric density structure of all models was perturbed with an
$m=2$ perturbation with different amplitude as they were introduced
into the hydrodynamics code.
%%; others were perturbed with
%%low-amplitude white noise fluctuations.
In the discussion that follows, the evolution time of individual
simulations has been normalized to the spin period at the rotation
axis of each star $T_c$, which varies significantly for different
models, but in terms of the characteristic dynamical time, $T_0$,
they were evolved for roughly the same times, except for model {\bf
J127}.  Table \ref{table:freq} summarizes all the relevant
frequencies and time scales for each of our models. In addition to
the time scales and frequencies defined earlier, $\tau_m$ is the measured e-folding
(growth) time for a given mode $m$, and $\omega_s$ is the angular
frequency at the equatorial surface.
We note that, for a given model, the eigenfrequencies quoted in Table \ref{table:freq}
have been measured at different evolutionary times. 
Each $\omega_m$ is measured at a time when the corresponding mode $m$
is dominant, and the corresponding corotation radius
is determined from the system's equatorial plane
rotational profile at that same time.

\subsection{Evolution of model J250}

This investigation began as an extension of Paper I (where
uniformly rotating neutron stars were studied) and our initial aim
was to study the secular bar-mode instability in differentially
rotating neutron stars. The first model we evolved was model {\bf
J250}, which was expected to be secularly unstable but dynamically
stable to the high $T/|W|$ bar-mode instability. Because the model
had a relatively large value of the parameter $A$,
it also was expected to be stable
to the low $T/|W|$ dynamical bar-mode instability identified by
\cite{SKE02}. Model {\bf J250} was evolved for many dynamical
times with an artificially enhanced GRR potential, in a manner
similar to the evolutions described in Paper I.

\begin{figure}[ht]
\epsscale{1.0} \plotone{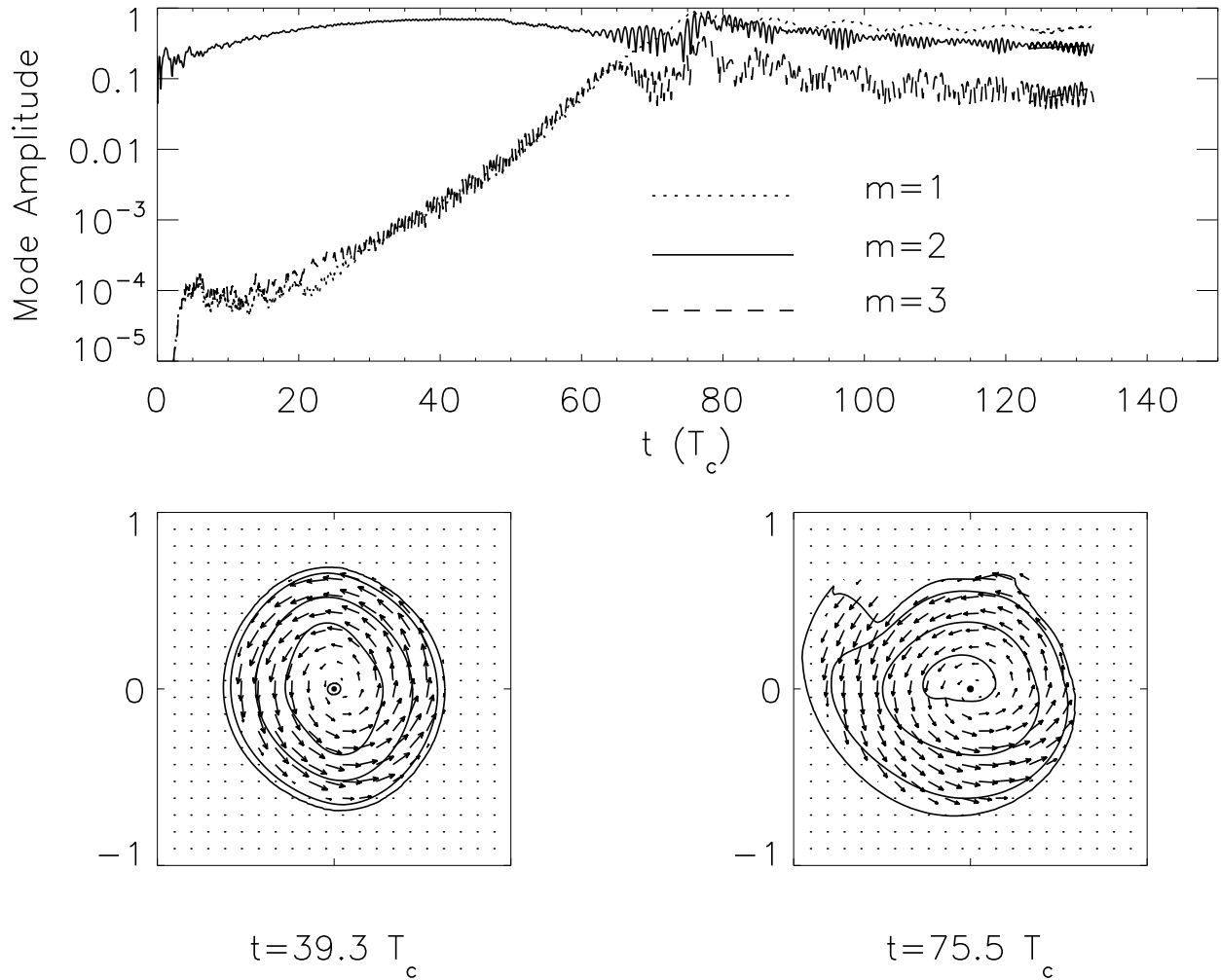}
%%\centerline{\includegraphics[width=3.5in]{allplots.j250.ps}}
%%\epsscale{1.0} \plotone{allplots.j250.ps}
\caption{The top panel shows $m=1$, $2$, and $3$ Fourier amplitudes
as a function of time $t$ in the equatorial plane of model {\bf
J250} when GRR was turned on; $t$ is normalized to the central spin
period $T_c$ of the differentially rotating star. 
The bottom panels show equatorial-plane
isodensity contours at two different evolutionary times, with
inertial-frame velocity vectors superposed; contour levels are (from
the innermost, outward) $\rho/\rho_\mathrm{max} = 0.9, 0.5, 0.1$,
and 0.01.
  \label{fig:mode250G}
}
\end{figure}

The top panel of Figure \ref{fig:mode250G} shows the
time-evolution of the amplitudes of different modes 
(averaged over the whole volume)
throughout this evolution. In the beginning, the $m=2$ bar-mode
grows exponentially on a timescale governed by the artificially
enhanced GRR potential; but it saturates after only a moderately
deformed bar structure has formed. This is consistent with the
results reported by \cite{SK04}.  However, on a considerably
longer time scale, an unexpected $m=1$ mode arises and
becomes the dominant mode. The density contour plots in the bottom
right-hand panel of Figure \ref{fig:mode250G} suggest that the
$m=1$ mode has a one-armed spiral structure similar to
the one discovered by \cite{CNLB01}. We note that, late in the
evolution, the amplitude of the $m=2$ mode increases when
the $m=1$ mode reaches the nonlinear regime. (This
feature is common to all of our models that are unstable to the
$m=1$ mode.) Measurements of $\omega_m$ for various modes
at the time when the $m=1$ mode is dominant confirms
that, for example, $\omega_2 \approx 2\omega_1 \approx 2.6$
%%the frequency of the $m=2$ mode equals that of the $m=1$ mode,
and suggests that, at late times, the $m=2$ mode is a
harmonic of the $m=1$ mode.

This discovery is unexpected in the sense that the one-armed, $m=1$
instability had previously appeared only in models with a softer EOS
($n \gtrsim 2.5$) and with a high degree of differential rotation,
i.e., $\omega_c/\omega_s \approx 10$
\citep{CNLB01,saijo:03,saijo:05}, where $\omega_s$ is the angular
frequency at the equatorial surface. In contrast, our model {\bf
J250} has an $n=1$ polytropic EOS and has only a moderate degree of
differential rotational with $\omega_c/\omega_s \approx 2$.

\begin{figure}[]
%%\centerline{
%%\includegraphics[width=3.5in]{allplots.j250.k0.ps}}
%%\epsscale{1.0} \plotone{allplots.j250.k0.ps} 
\epsscale{1.0} \plotone{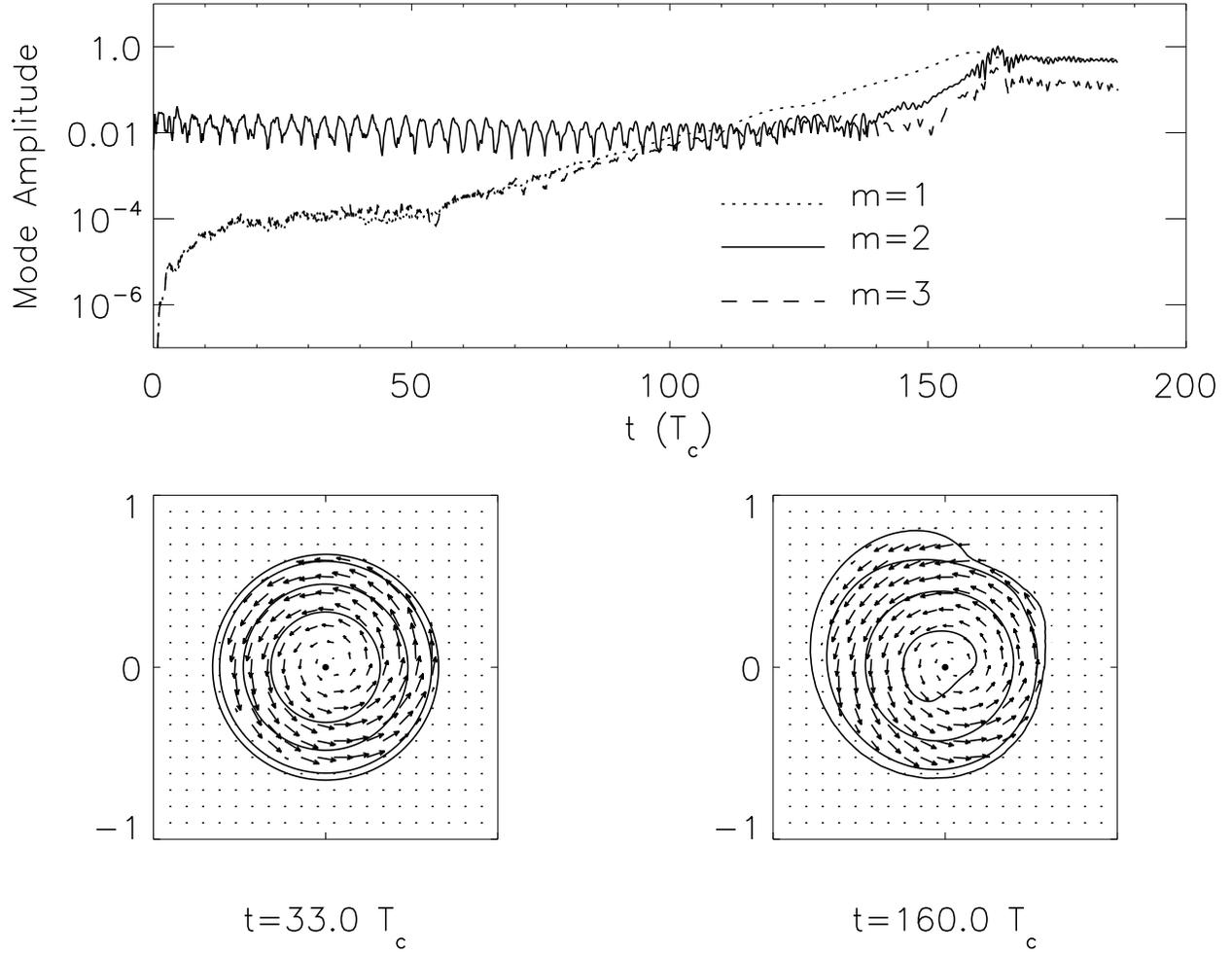} 
\caption{Same as Figure
\ref{fig:mode250G}, but for model {\bf J250} when GRR forces are
turned off. \label{fig:mode250} }
\end{figure}

\begin{figure}[]
%%\centerline{
%%\includegraphics[width=3.5in]{drho.j250.ps}}
%%\epsscale{1.0} \plotone{drho.j250.ps}
\epsscale{1.0} \plotone{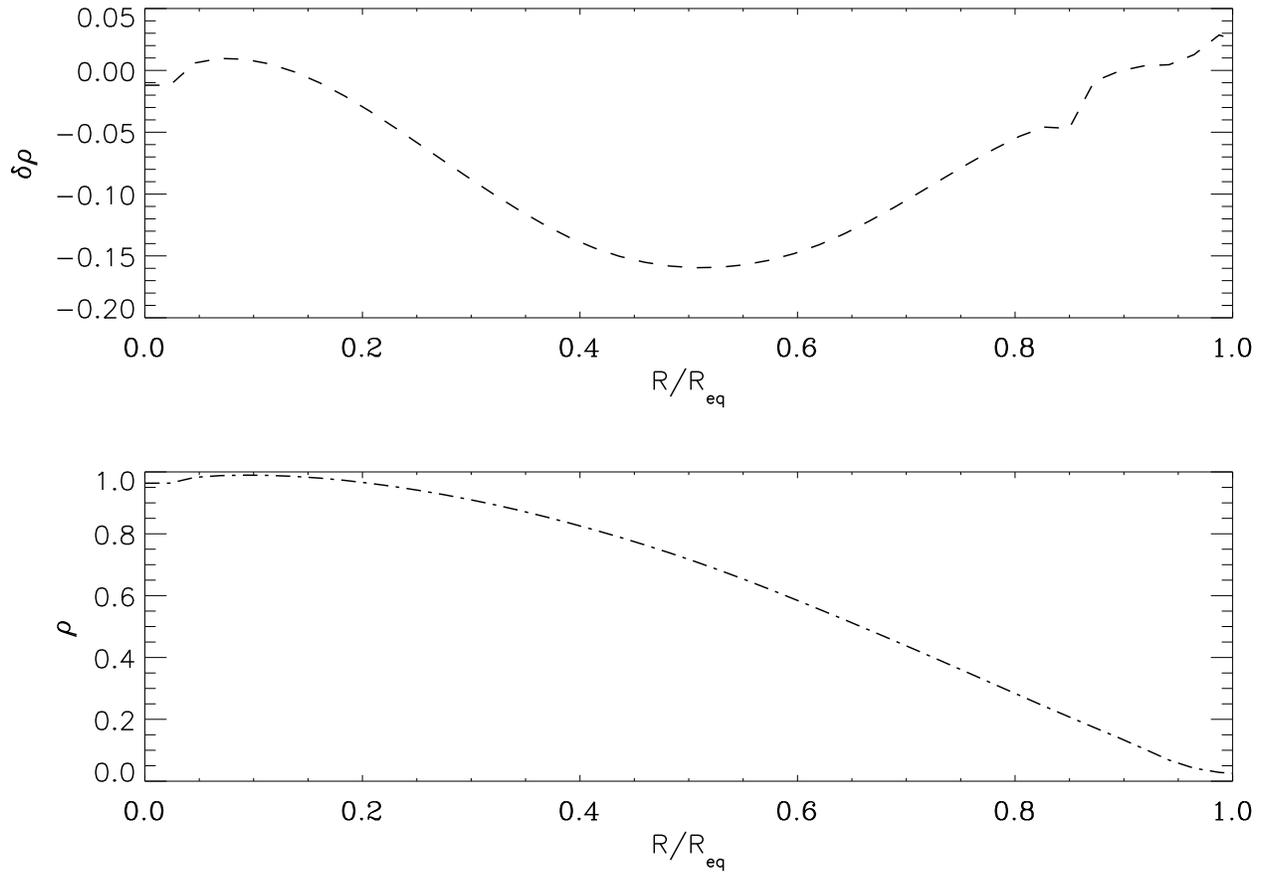}
\caption{The top and bottom panels show, respectively, $\delta\rho(R)$
and $\rho(R)$ for model {\bf J250} in the equatorial plane 
when the $m=1$ mode is dominant.
\label{fig:drhoj250} }
\end{figure}

\begin{figure}[]
%%\centerline{
%%\includegraphics[width=3.5in]{omega.j250.ps}}
%%\vskip.3cm
%%\epsscale{1.0} \plotone{omega.j250.ps} 
\epsscale{1.0} \plotone{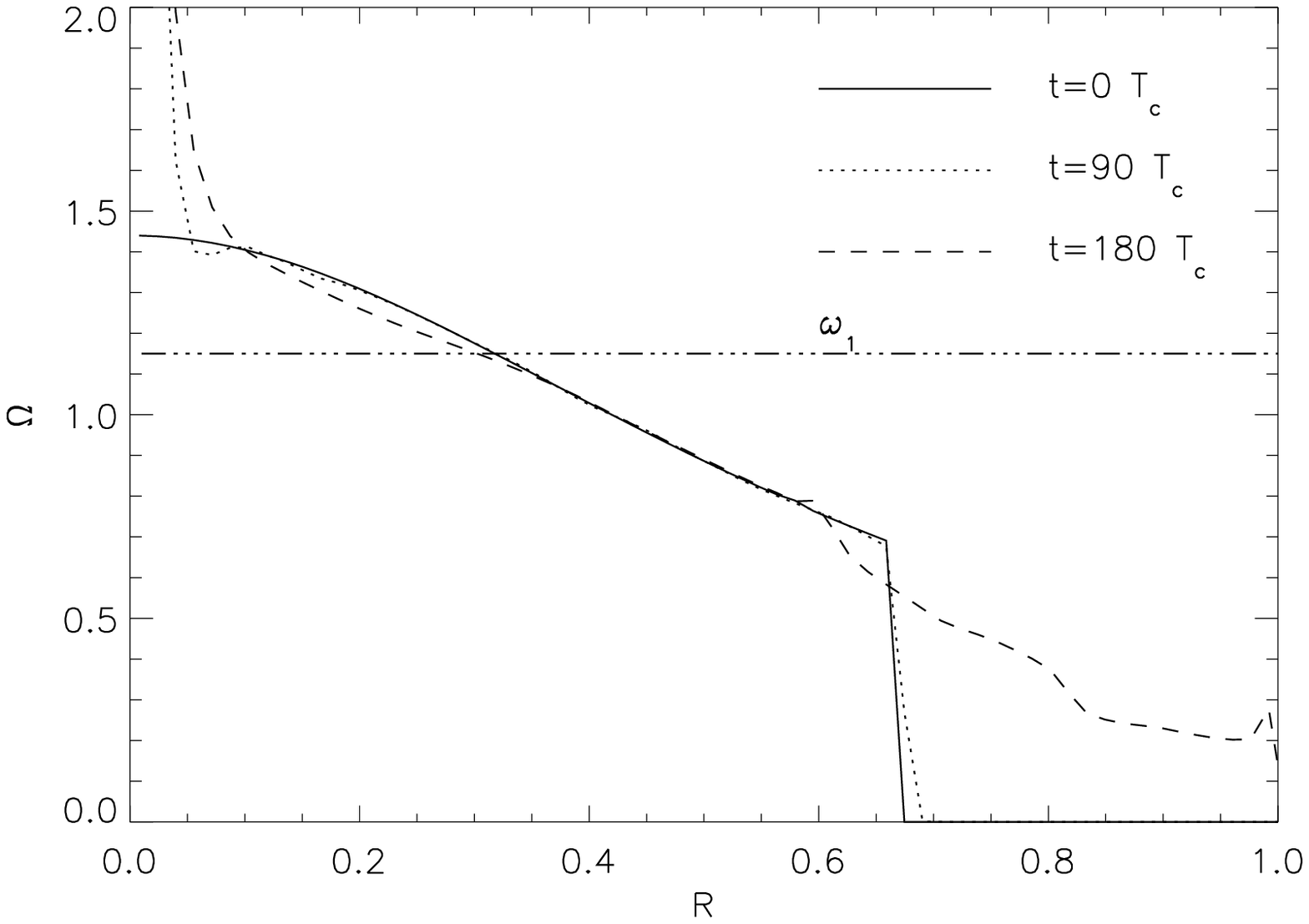} 
\caption{The rotational
profile $\Omega(R)$ of model {\bf J250} is shown at three different
evolutionary times. The dash-dotted horizontal line identifies the
measured eigenfrequency $\omega_1$ of the unstable m=1 mode.
\label{fig:corotj250} }
\end{figure}

In order to test whether or not this instability was caused by the
artificially enhanced GRR forces, a second evolution was performed
on model {\bf J250} with GRR turned off. Figure \ref{fig:mode250}
shows the time-evolution of the amplitudes of different modes for
this second evolution.  As expected, the $m=2$ bar-mode was not
unstable in the absence of GRR forces, but surprisingly the $m=1$
mode remained unstable, becoming the dominant Fourier component at
time $\gtrsim 100 ~T_c$. Compared to models that had a softer EOS
\citep{CNLB01,saijo:03} in which the $m=1$ mode grew on a time scale
of a few tens of $T_c$, the growth time here is much longer but
nevertheless still dynamical. (We note that the growth time in this
pure hydrodynamical evolution is also longer than that from the
previous run, in which GRR was turned on. This is probably because
GRR tends to increase the degree of differential rotation, hence
influence the growth rate of the $m=1$ mode.)
%%Therefore the instability is sufficient to surpass the secular bar-mode
%%instability even from the beginning.
The top panel of Figure \ref{fig:drhoj250} shows the radial
eigenfunction of the $m=1$ mode obtained by subtracting the initial
axisymmetric background density (bottom panel of Figure
\ref{fig:drhoj250}). Its behavior is qualitatively similar to that
derived from a linear stability analysis by \cite{saijo:05} (see
their Figure 9).

The appearance of the one-armed spiral instability in our models
with a stiff EOS seems to suggest that its stability criterion is
independent of compressibility. Now the question is what mechanism
is responsible for its growth? According to \cite{watts:04},
\cite{saijo:05}, and related studies of disks and tori
\citep{PP84,GGN86,NGG87}, corotation points may be responsible for
the amplification of some unstable modes. But this mechanism of
amplification can only operate if there is at least one radial
location in the model where the angular frequency of orbiting fluid
elements resonates with the eigenfrequency of the identified special mode. 
A straightforward check of the radial rotational profiles $\Omega(R)$
of model {\bf J250} (see Figure \ref{fig:corotj250}) shows that a
corotation radius is present at $R \approx 0.32$ for the unstable
$m=1$ Fourier mode with an eigenfrequency $\omega_1 \approx 1.2$.
(This measurement of the eigenfrequency was done for the pure
hydrodynamical simulation, in which GRR was turned off.)

However, the result from one model is not sufficient for us to
draw a definite conclusion.  To further clarify the relationship
between the instability and corotation points, we performed a
number of simulations on different polytropic neutron star models. 
In these
additional simulations, we will concentrate only on hydrodynamic
simulations with GRR turned off, because GRR-related instabilities
generally would set in on a much longer time scale and therefore
are irrelevant here.

\subsection{Evolution of model J133}

So far we have identified a corotation point associated with the
one-armed spiral instability in model {\bf J250}. It would be
interesting to find out if the instability disappears when there
is no such corotation point. In order to test this, we chose to
evolve model {\bf J133}. This model is rotating slowly enough so
that the central angular frequency is well below the
eigenfrequency of the $m=1$ mode that arose in model {\bf
J250}. By removing the corotation points of the $m=1$
mode inside model {\bf J133}, it was hoped that the one-armed
instability would not show up in this evolution. To eliminate the
possible influence of the degree of differential rotation, we kept
the value of the parameter $A$ the same as in model
{\bf J250}, but increased the polar radius $R_p$ to slow down the
rotation of the star. Note that the eigenfrequency of the
$m=1$ mode for this model is actually
unknown\footnote{Our studies do not include a method for
performing a complete eigenmode analysis of our selected,
initially axisymmetric models. Only modes that are unstable
and that grow on a timescale short compared to our evolution
times can be identified using our present nonlinear
hydrodynamic tools.}, but we estimate it would not
change dramatically compared to that of model {\bf J250}. (In
fact, as shown by later simulations, $\omega_1$ in model {\bf
J250} and {\bf J127} are very close to each other; see Table
\ref{table:freq}.)

\begin{figure}[]
%%\centerline{
%%\includegraphics[width=3.5in]{allplots.j133.ps}}
%%\epsscale{1.0} \plotone{allplots.j133.ps}
\epsscale{1.0} \plotone{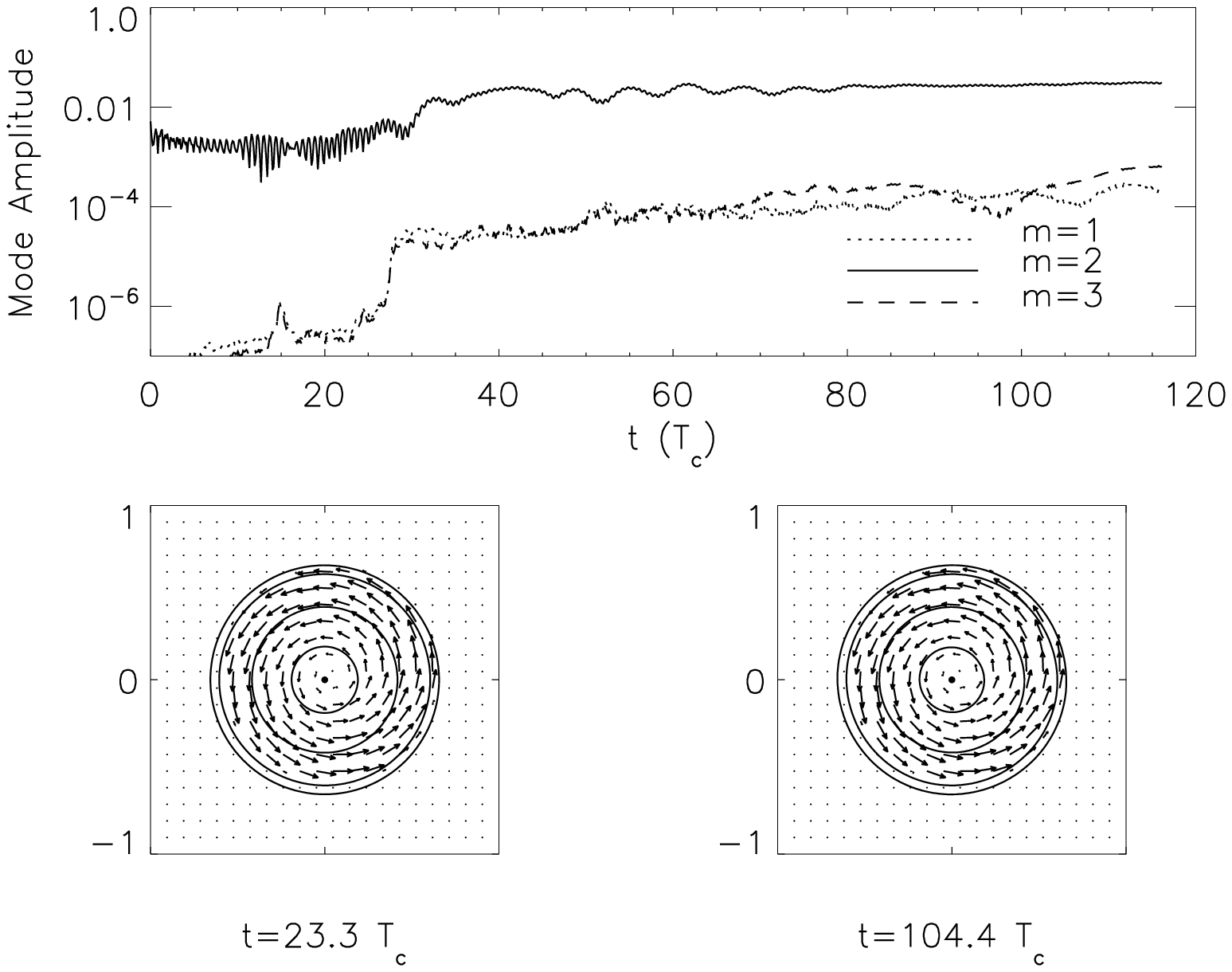}
\caption{Same as Fig. \ref{fig:mode250G}, but for model 
 {\bf J133} when GRR forces are turned off. \label{fig:mode133} }
\end{figure}

As shown in the top panel of Figure \ref{fig:mode133}, the $m=1$
Fourier amplitude $|C_1|$ does not grow into the nonlinear regime in
this case. It develops somewhat, but only hangs on at a very small
amplitude.  On the other hand, the $m=2$ mode seems to
gain some strength in the beginning of the evolution, but it
quickly saturates at a very low amplitude and never enters into
the nonlinear regime. The density contours shown in the bottom
panels of Figure \ref{fig:mode133} also display no indication of
nonaxisymmetric structure. The lack of development of the
$m=1$ spiral mode in this simulation supports our
conjecture that corotation points are directly responsible for the
excitation of the mode.

\subsection{Evolution of model J127 }

Although the evolution of model {\bf J133} makes our argument
about corotation points plausible, the fact that its $T/|W|$ value
is significantly lower than model {\bf J250} raised the following
question: is it the low $T/|W|$ value that suppressed the
instability? To answer this question and as a further check of our
conjecture, model {\bf J127} was evolved in the same fashion as
model {\bf J133}. This model has a $T/|W|$ value that is similar
to that of model {\bf J133}, but its degree of differential
rotation was made very high by reducing the value of $A$.  With
this steep rotational profile, we hoped that a corotation point
for the $m=1$ mode would exist within the star and,
hence, trigger the instability.

\begin{figure}[]
%%\epsscale{1.0} \plotone{allplots.j127.ps}
\epsscale{1.0} \plotone{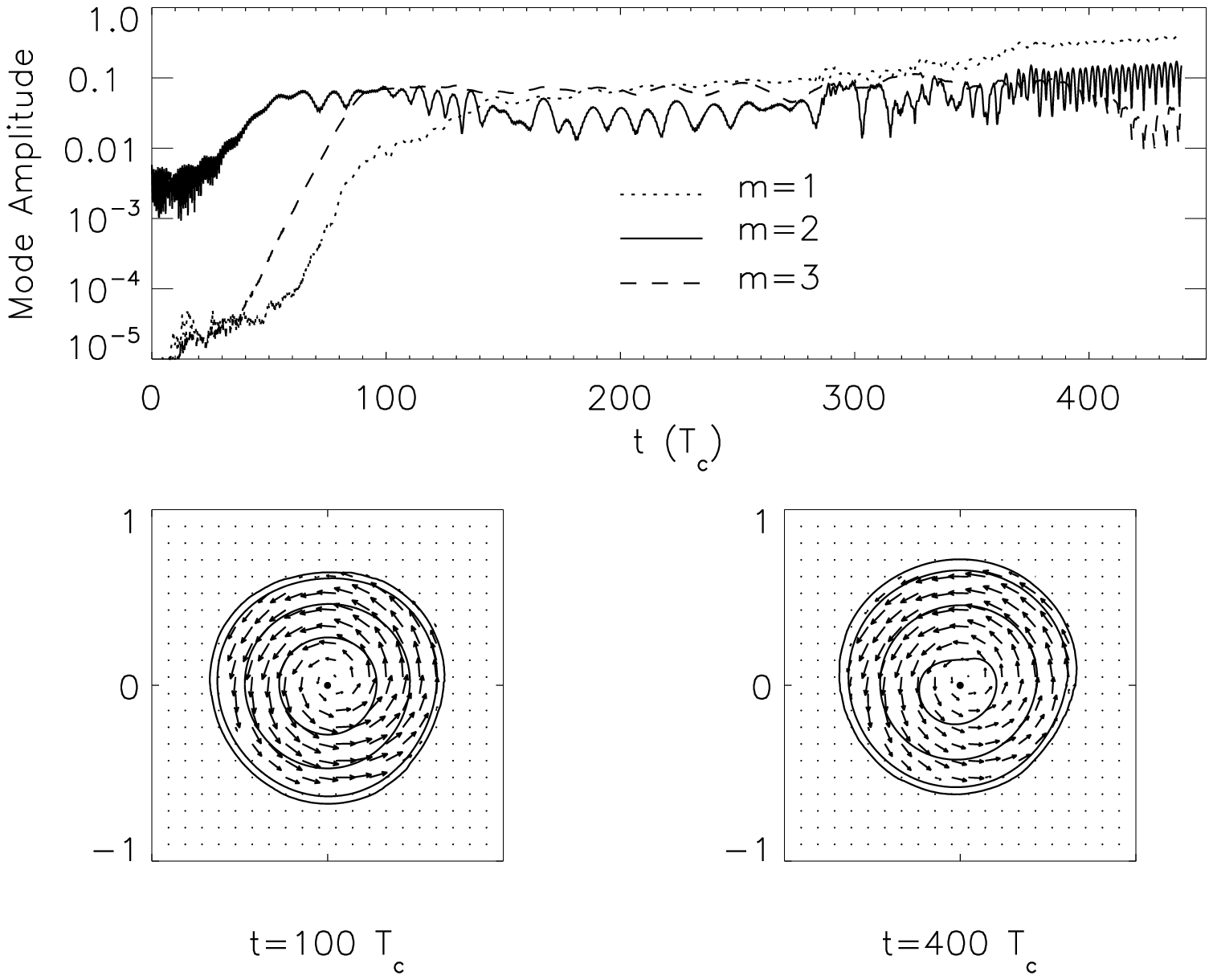}
\caption{The same as Figure \ref{fig:mode250G}, but for model {\bf
J127} with GRR turned off. \label{fig:mode127} }
\end{figure}

The final outcome of this evolution is perhaps the most striking
and interesting one.  As shown in the top panel of Figure
\ref{fig:mode127}, the evolution can be divided into three stages:
(1) In the first $\sim 50 ~T_c$, an $m=2$ mode grows
exponentially, but it saturates very quickly;
(2) $\sim 30 ~T_c$ later, an unexpected $m=3$
mode comes into play, and becomes dominant for quite a long time.
The pear-shaped $m=3$ density distortion is visible in
the bottom left-hand panel of Figure \ref{fig:mode127}; (3) after
$\approx 300 T_c$, the originally expected $m=1$ mode
catches up and wins over.

\begin{figure}[]
%%\centerline{
%%\includegraphics[width=3.5in]{omega.j127.ps}}
%%\epsscale{1.0} \plotone{omega.j127.ps}
\epsscale{1.0} \plotone{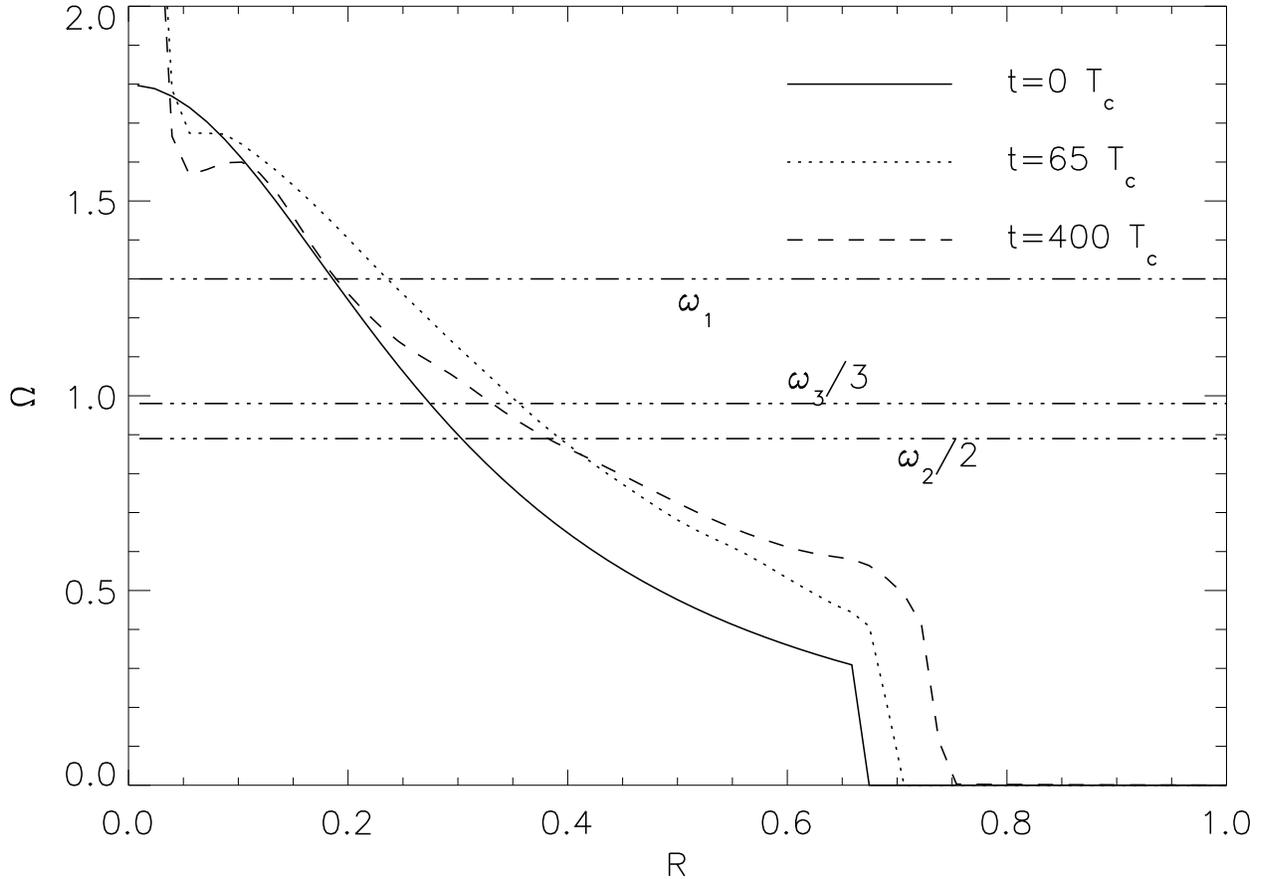}
\caption{The angular velocity profile $\Omega(R)$ is shown for model
{\bf J127} at three different evolutionary times. Dash-dotted
horizontal lines denote the measured eigenfrequencies $\omega_1,
\omega_2$ and $\omega_3$ of various unstable modes.
$\omega_2/2$ is measured at $\sim 50$ $T_c$,
$\omega_3/3$ is measured at $\sim 80$ $T_c$,
and $\omega_1$ is measured at $\sim 400$ $T_c$.
To determine a corotation radius,
we used a rotational profile that is around at the same time
when each frequency is measured.
\label{fig:corotj127} }
\end{figure}

Given previous studies of the low $T/|W|$ bar-mode instability
\citep{SKE02}, the appearance of the $m=2$ mode here is
not surprising. However, the appearance of an $m=3$ mode
in axisymmetric equilibrium stellar models with intermediate
$T/|W|$ values is totally unexpected. A plot of the rotational
profile of model {\bf J127} is shown in Figure \ref{fig:corotj127}
and reveals
the existence of corotation radii for all three of these unstable
modes. These results are fully consistent with the linear analysis
presented by \cite{watts:04}: Corotation instabilities in stellar
models are not limited to $m=1$ and $m=2$ modes.
Other modes may become unstable as long as their corotation points
exist inside the star.
In this simulation, we also analyzed the behavior of the $m=4$ ($5$) 
Fourier component of the density distribution.
It appears to follow the behavior of the $m=2$ ($3$) mode, 
but it saturates at an amplitude that is two orders
of magnitude lower. This suggests that the $m=4$ ($5$) ``mode" 
is actually a harmonic of $m=2$ ($3$).  
There is no indication of the independent development of
either an $m=4$ or $5$ mode.
We suspect that the reason for the lack of independent m=4
and 5 modes is that their corotation radii fall outside the star.

\begin{figure}[]
%%\centerline{
%%\includegraphics[width=3.5in]{strain.j127.ps}}
%%\epsscale{1.0} \plotone{strain.j127.ps} 
\epsscale{1.0} \plotone{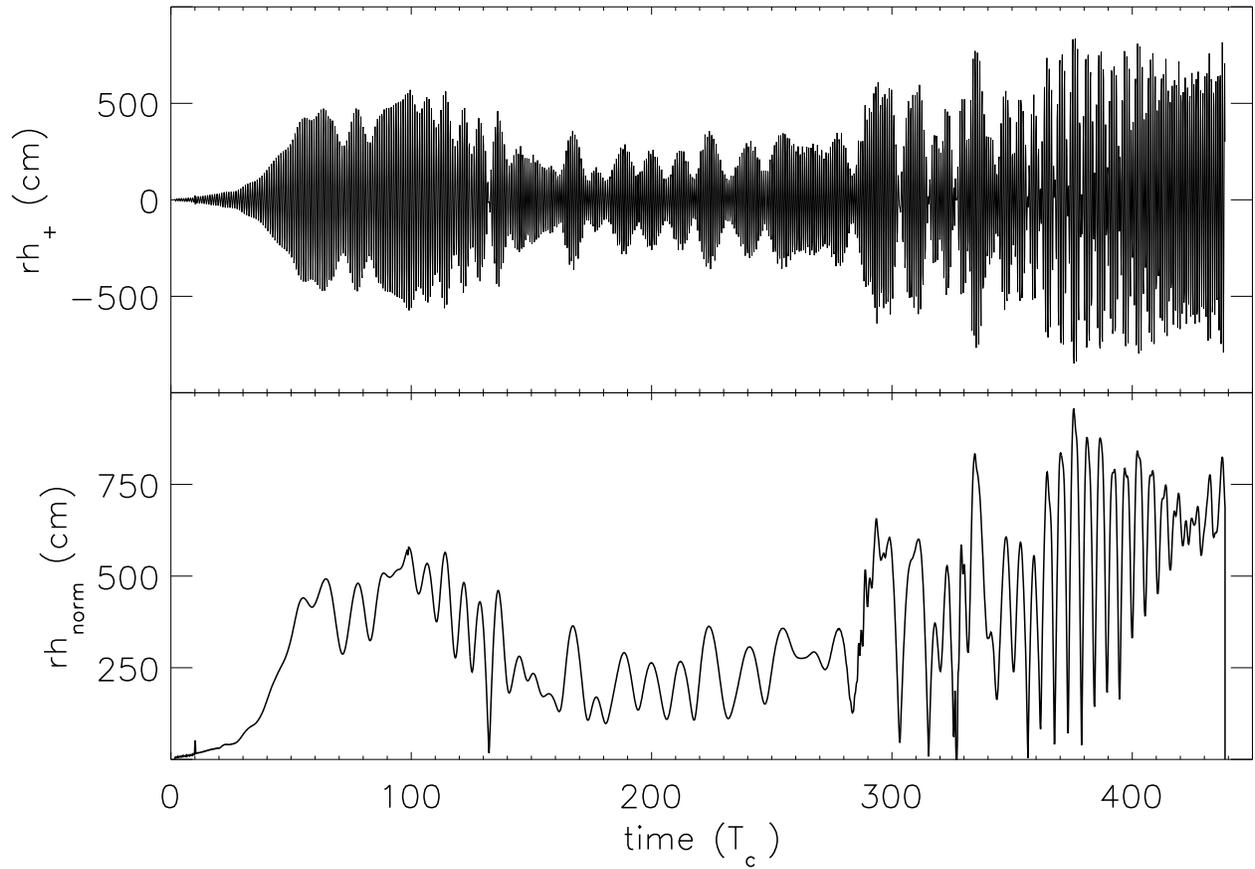} 
\caption{The quadrupole
gravitational-wave strain that would be radiated from model {\bf
J127} as viewed by an observer looking down the $Z$ (rotation) axis.
\label{fig:gw_j127} }
\end{figure}

Because instabilities associated with different modes may arise in
stars with strong differential rotation, the gravitational-wave
signals from such an event would exhibit multiple characteristics.
Figure \ref{fig:gw_j127} shows the computed quadrupole
gravitational-wave strain as viewed by an observer located along the
+Z axis and assuming the neutron star has a mass of $1.4~M_{\odot}$
and an equatorial radius of 12.5 km. The peak signal strength is
actually stronger than the bounce signal expected from the
axisymmetric core collapse of massive stars, which has been shown to
produce signals with $rh_+ \approx 300$ cm at a frequency $f\approx
400$ Hz \citep{ott:04}. The signal also has some distinct features.
There is more than one localized burst, arising from different modes
peaking at different times and in different frequency bands: the
first, centered at $t/T_c \sim 65$, corresponds to  $rh_+$ $\sim$
500 cm at a frequency $f \approx 1800$ Hz emitted by the $m=2$ mode;
%% ($\omega_2 \approx 1800$ Hz);
the second, centered at $t/T_c \sim 100$, corresponds to $rh_+\sim
600$ cm at a frequency $f \approx 1900$ Hz associated with the $m=3$
mode;
%%($\omega_3 \approx 2800$ Hz);
the third, centered at $t/T_c \sim 375$, corresponds to a $rh_+ \sim
900$ cm at frequency $f \approx 2600$ Hz emitted by the $m=2$
harmonic of the $m=1$ spiral mode.
%%($\omega_1 \approx 1200$ Hz).

\begin{figure}[]
\epsscale{1.0} \plotone{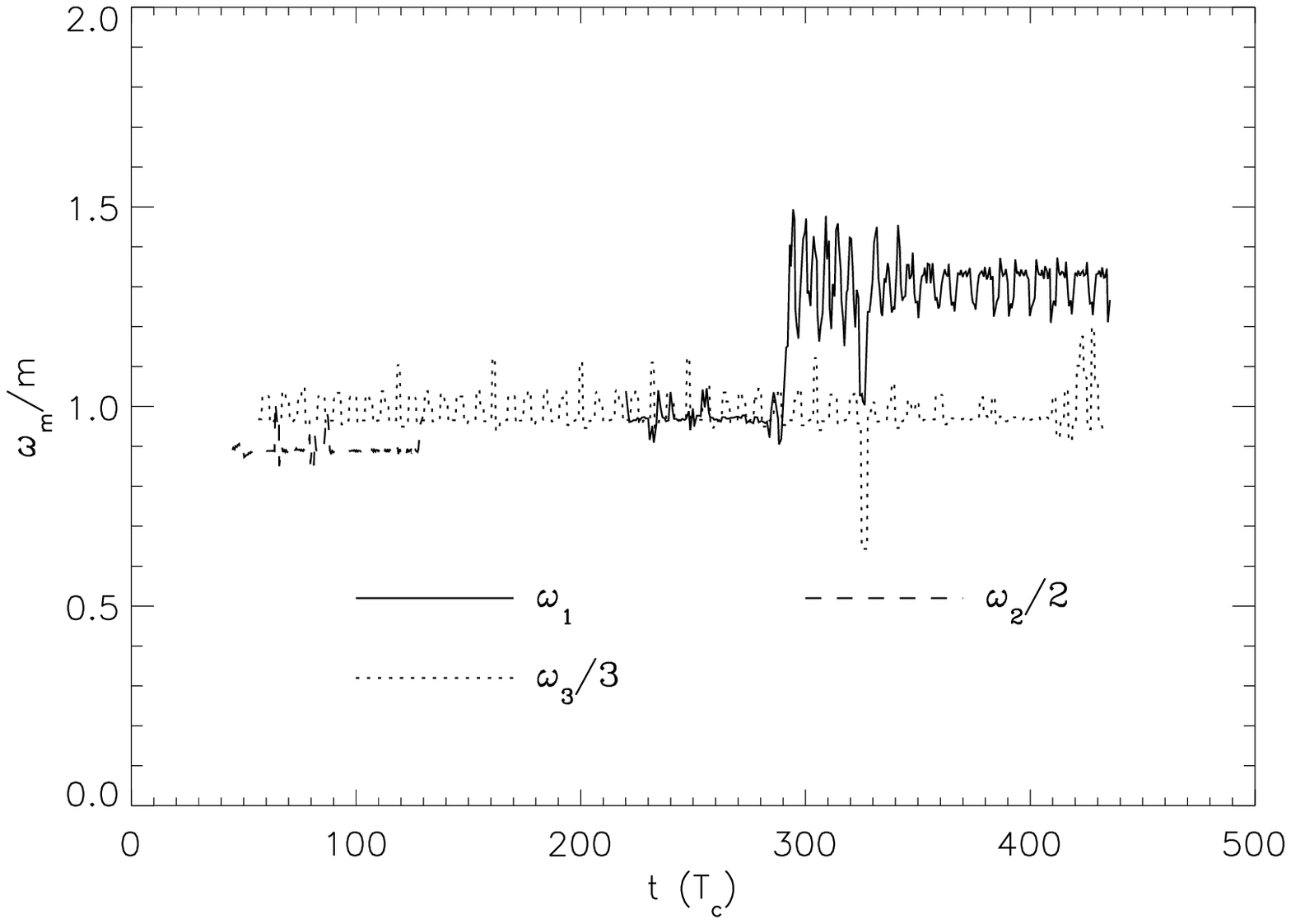}
\caption{The pattern frequencies of different modes is displayed
as a function of time for model {\bf J127}.
The frequency $\omega_2/2$ is measured between $40$ and $130$ $T_c$,
$\omega_3/3$ is measured between $50$ and $430$ $T_c$,
and $\omega_1$ is measured between $220$ and $430$ $T_c$.
\label{fig:freq_j127} }
\end{figure}

Figure \ref{fig:freq_j127} shows the pattern frequencies of all 
the dominating unstable modes as a function of time.
To make the graph clean, the frequencies of the $m=1$ and $2$ modes
are only shown in the time intervals when they dominate,
and all the data at times $t < 40 T_c$ have been omitted 
because they appear to be noisy due to the lack of
coherent modes.
It is observed that the pattern frequencies of all three modes
-- $\omega_1, \omega_2/2$, and $\omega_3/3$ -- are fairly constant 
throughout their peroid of dominance and 
the uncertainty in these measurements  
is $\lesssim 10 \%$.
Hence, the relative location of the corotation radii $R_{\mathrm{cor}}$
for these various modes does not vary with time, 
unless a change in the rotational profile of the star causes
$R_{\mathrm{cor}}$ to migrate.

Based upon the above hydrodynamic simulations, we can now draw the
following conclusions: (1) The $m=1$ instability can be
turned on or off by controlling the rotational profile of the star
to allow or remove the existence of a corotation point. (2) The
instability is not limited to the $m=1$ mode; other modes
may also arise, as long as their corotation points exist inside
the star.

\subsection{Evolution of model J068}

According to \cite{SKE02, SKE03}, models with very small values of
the rotation-profile parameter $A$ are subject to a low $T/|W|$
dynamical bar-mode instability. Since the rotational profiles for
these types of models are quite steep, it occurred to us that they
may also allow corotation points for various modes within their
structure. It also seemed likely that even one individual stellar
model might be subject to instabilities associated not only with
the $m=2$ bar-mode, but also other modes. In order to
test this idea, we evolved model {\bf J068} with
$A/R_{eq}=0.44$ ($\omega_c/\omega_s\approx 6$), which
falls in the instability region where the low $T/|W|$ dynamical
bar-mode instability would be expected to set in.

\begin{figure}[]
\epsscale{1.0} \plotone{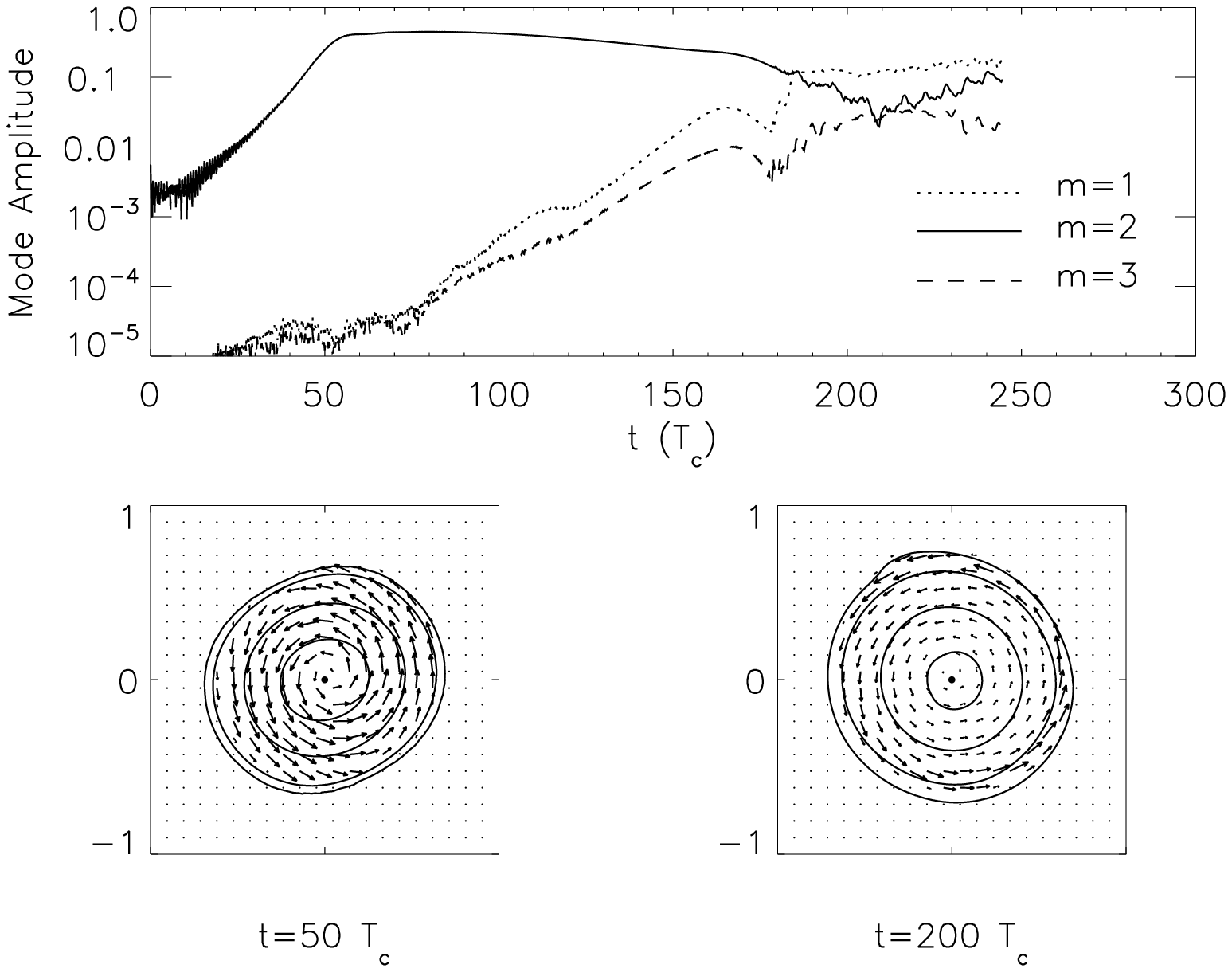}
\caption{Same as Figure \ref{fig:mode250G}, but for model {\bf J068}
when GRR is turned off. \label{fig:mode068} }
\end{figure}

The results of this simulation support our suspicion that an
individual model with a very steep rotational profile may be
subject to multiple instabilities associated with different
azimuthal modes. As shown in the top panel of Figure
\ref{fig:mode068}, at the beginning of this simulation the
$m=2$ bar-mode grew very quickly, became nonlinear, and
dominated the evolution for quite a while, which is consistent
with the results presented by \cite{SKE02}. In addition, however,
we found that the underlying $m=1$ and $3$ modes also
grew simultaneously. These odd-numbered modes had growth rates
rather slower than the $m=2$ mode, but this did not keep
the $m=1$ mode from finally winning over all the other
modes, including the bar-mode, and manifesting itself as the
dominant mode, just as it did in our earlier simulations. Notice
that, again, the amplitude of the $m=2$ mode increases when
the $m=1$ mode becomes dominant at late times, which
suggests that the $m=2$ mode is a harmonic of the
$m=1$ mode during this period.

\begin{figure}[]
\epsscale{1.0} \plotone{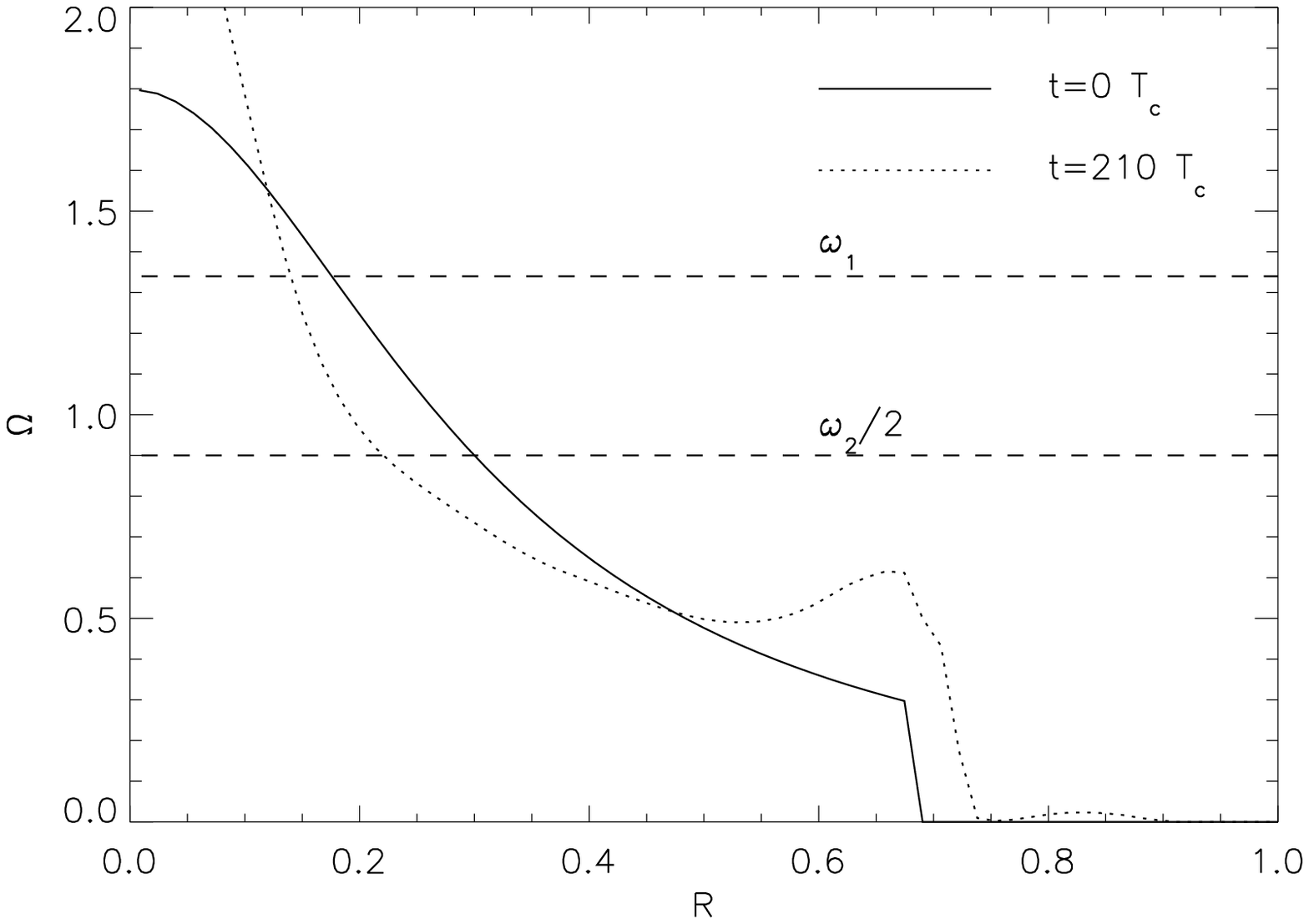} 
\caption{The angular velocity
profile $\Omega(R)$ of model {\bf J068} is plotted at two different
times during its evolution. Dashed horizontal lines identify
measured eigenfrequencies of the unstable $m=1$ and $m=2$ modes.
\label{fig:corotj068} }
\end{figure}

The rotational profile of model {\bf J068}, shown in
Figure \ref{fig:corotj068}, reveals corotation points
for both $m=1$ and $2$ modes.
(We could not accurately measure the eigenfrequency of
the $m=3$ mode because it never became the
dominant mode in the evolution time followed
by our simulation.)

%%\subsection{Evolution of model V254}

\section{Summary and Discussion}

We have found that a one-armed spiral instability arises in rotating
$n=1$ ``polytropic" neutron stars with moderate and high degrees of
differential rotation.   This is in contrast to the work of
\cite{saijo:03}, which has suggested that such instabilities can
arise only in stars with a softer EOS. As has been suggested by
\cite{saijo:05}, the instability seems to be
associated with the existence of a corotation point of the $m=1$
mode that lies inside the star. Given previous results for models
with a softer EOS, we conclude that the compressibility and $T/|W|$
value of the star are not directly responsible for triggering the
instability.
%%but may affect the growth rate.

We have discovered that, in addition to the $m=1$ mode, other
higher-order modes also are allowed to come into play within a
single model as long as corotation points for these modes also exist
within the star.  This can happen as long as the rotational profile
is sufficiently steep.
%%Furthermore, in these corotation instabilities,
%%each unstable mode has different growth rate at different $T/|W|$.
%%A rough estimation can be drawn considering previous studies:
%%the $m=2$ mode seems to have the fastest growth rate
%%in $T/|W|$ ranges from 0.01 to around 0.1 \citep{SKE03},
%%the $m=1$ mode seems to grow faster
%%in $T/|W|$ ranges from around 0.13 to 0.25,
%%the $m=3$ mode may set in at intermediate $T/|W|$ values.
In our $n=1$ polytropic neutron star models, these modes grow
relatively slowly compared to previous results for softer EOSs
\citep{CNLB01,saijo:03}, but nevertheless they appear to be
dynamically unstable.
%%still dynamical; hence, they are sufficient
%%to surpass the secular bar-mode instability. Therefore, for stars
%%with an extremely steep rotational profile, which may allow the
%%corotation points for various modes to fall within the star, the
%%instability criterion is not unique because various modes might set
%%in due to these corotation related instabilities.
Such modes are likely to develop in newly born neutron stars because
young neutron stars are expected to have strong differential
rotation \citep{ott:05b, AW05}. Also, because they develop in low
$T/|W|$ configurations on a dynamical time scale, these modes are
likely to be more important than the often-discussed secular
bar-mode instability.  This will have important consequences for the
detection of gravitational-waves by the Laser Interferometer
Gravitational-wave Observatory (LIGO), because gravitational-waves
generated in such systems would not be monochromatic; our
simulations indicate that signals with different characteristic
frequencies are likely to peak at different times.

We suspect that these corotation-related instabilities have not been
identified in previous studies of models that have a stiff EOS
because earlier studies have been focused on a search for, or the
analysis of, instabilities that develop on a very short time scale.
Most previous simulations have been followed through only a few tens
of $T_c$, which is not a long enough time to permit significant
amplification of many of the modes we have identified in this study.
Only in this study -- which was initially aimed at examining the
secular bar-mode instability in stars with differential rotation --
have models been evolved long enough to reveal the corotation
instabilities that grow on a longer time scale for this stiff EOS
case.

\begin{figure}[]
%%\centerline{
%%\includegraphics[width=3.5in]{vortensity.j068.ps}}
%%\epsscale{1.0} \plotone{vortensity.j068.ps} 
\epsscale{1.0} \plotone{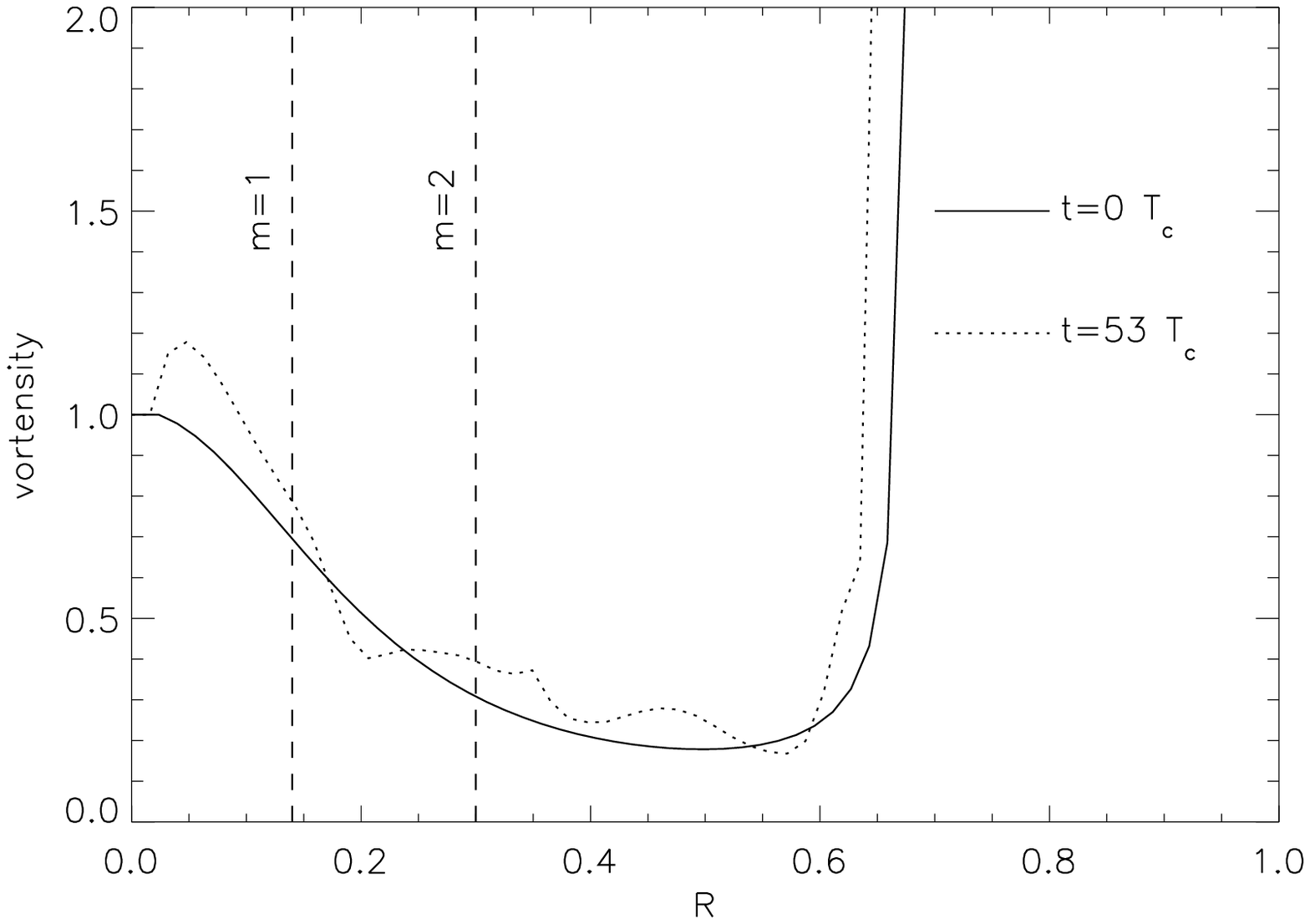} 
\caption{The radial
vortensity profile is shown for model {\bf J068} initially (solid
line) and at a time $t = 53 T_c$ (dotted line) during its evolution.
The horizontal axis is the cylindrical radius, the vertical axis is
the vortensity normalized to the value on the $Z$ axis. The two
vertical dashed lines locate the corotation radii for $m=1$ (left)
and $m=2$ (right) modes. There seems to be a correlation between the
growth rate and the location of corotation radii inside the
vortensity well: the $m=2$ mode with a ``deeper" corotation radius
in the initial vortensity well was the first unstable mode to arise
in the simulation, whereas the $m=1$ mode whose corotation radius
lies at a ``shallower'' location in the initial vortensity well
developed later in the evolution. \label{fig:vort_j068} }
\end{figure}

\begin{figure}[]
%%\epsscale{1.0} \plotone{vortensity.j127.ps} 
\epsscale{1.0} \plotone{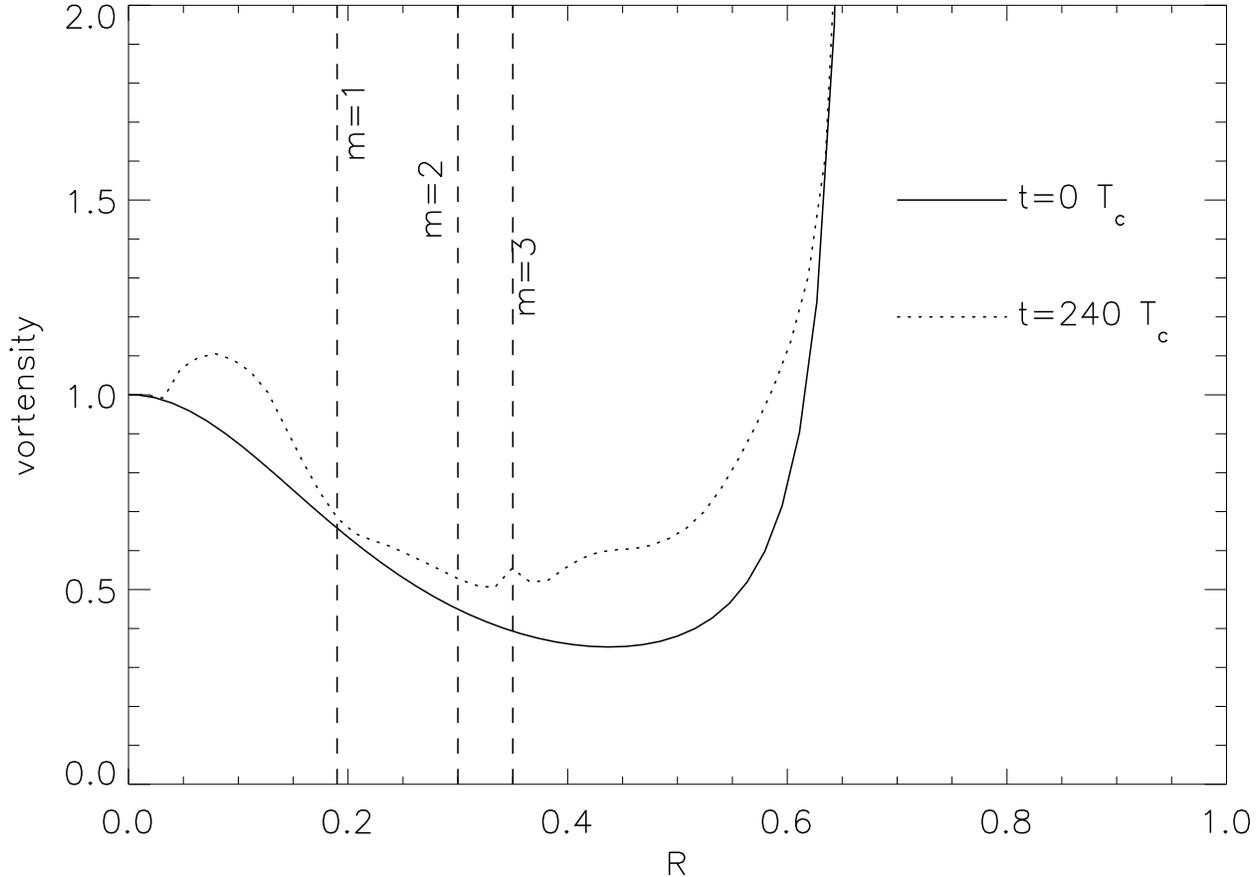} 
\caption{The radial
vortensity profile is shown for model {\bf J127} initially (solid
line) and at a time $t = 240 T_c$ (dotted line) during its
evolution. The horizontal axis is the cylindrical radius, the
vertical axis is the vortensity normalized to the value on the $Z$
axis. The three vertical dashed lines locate the corotation radii
for $m=1$ (left), $m=2$ (middle), and $m=3$ (right) modes. 
\label{fig:vort_j127} }
\end{figure}

Based on theoretical analyses of differentially rotating
configurations that have been presented by others, we suspect that,
in addition to the existence of corotation points, a resonant cavity
is required in order to drive corotation instabilities to very large
amplitude. In particular, \cite{LLCN99} have suggested that a local
minimum in the radial vortensity profile can serve as a resonant
cavity to trap waves and, hence, trigger a ``Rossby wave''
instability. Figures \ref{fig:vort_j068} and \ref{fig:vort_j127}
display the radial vortensity profiles of our models {\bf J068} and
{\bf J127} at different times. (The vortensity plotted here is the
ratio between vorticity and volume density, but the same radial
profile is preserved even if we define vortensity as the ratio
between vorticity and surface density.) As these figures show, a
fairly wide dip encompasses the corotation radii of all of the modes
that we have found to be unstable in these models. We note that such
a dip in the radial vortensity profile is a common feature for all
of our unstable models. We suspect that waves are trapped inside
these vortensity cavities and are continously amplified through
multiple passages across their respective corotation radii. Because
an off-axis density maximum alone will generate a vortensity minimum
and hence form a resonant cavity that can trap waves inside, it is
perhaps not surprising that toroidal structures frequently have been
found to be unstable to a one-armed, spiral instability
\citep{saijo:03}. However, our new results show that a toroidal
structure is not necessary to generate a vortensity minimum; strong
differential rotation alone can create a vortensity minimum, as
indicated for example by our model {\bf J068} which does not have a
toroidal structure.

We have also noticed that an unstable mode whose corotation radius
falls at a position that is deeper in the vortensity well 
%%develops earlier in any given simulation.  
has a higher growth rate in any given simulation.  
Figures \ref{fig:vort_j068} and
\ref{fig:vort_j127}, in conjunction with the data provided in Table
\ref{table:freq}, reveal this correlation between the growth rate of
each mode and the depth of its corotation radius inside the
vortensity well. This correlation is consistent with the results
reported by \cite{LFLC00} from a study of the Rossby-wave
instability in differentially rotating thin disks. 
They discovered that the
growth rate of the Rossby-wave instability was proportional to the
amplitude of density bumps/jumps that were introduced into their
models, in other words, proportional to the depth of the vortensity
wells that were created by such density jumps.

According to \cite{LFLC00}, requiring a minimum to be present in a
star's radial  vortensity profile before the star can become
susceptible to a corotation instability is a generalization of the
Rayleigh inflexion point theorem, which states that a necessary
condition for an instability of this type to develop is that there
be an inflection point in the velocity profile \citep{DR81}. Given
the j-constant-like rotation law (Eq.~\ref{EQ:jconst}) adopted in
our (as well as previous) studies, the condition $\partial^2
v_{\phi}/\partial^2 R=0$ can be evaluated analytically to show that
a velocity inflection point arises at $R=A/\sqrt{3}$.  Hence, for a
sufficiently shallow rotational profile having $A>\sqrt{3}
R_\mathrm{eq}$, the required inflection point does not exist inside
the star. This may also explain why low $T/|W|$ instabilities only
show up in stars with strong differential rotation (smaller values
of $A$).

\begin{figure}[]
%%\centerline{
%%\includegraphics[width=3.5in]{phase.j250.ps}}
%%\epsscale{1.0} \plotone{phase.j250.ps} 
\epsscale{1.0} \plotone{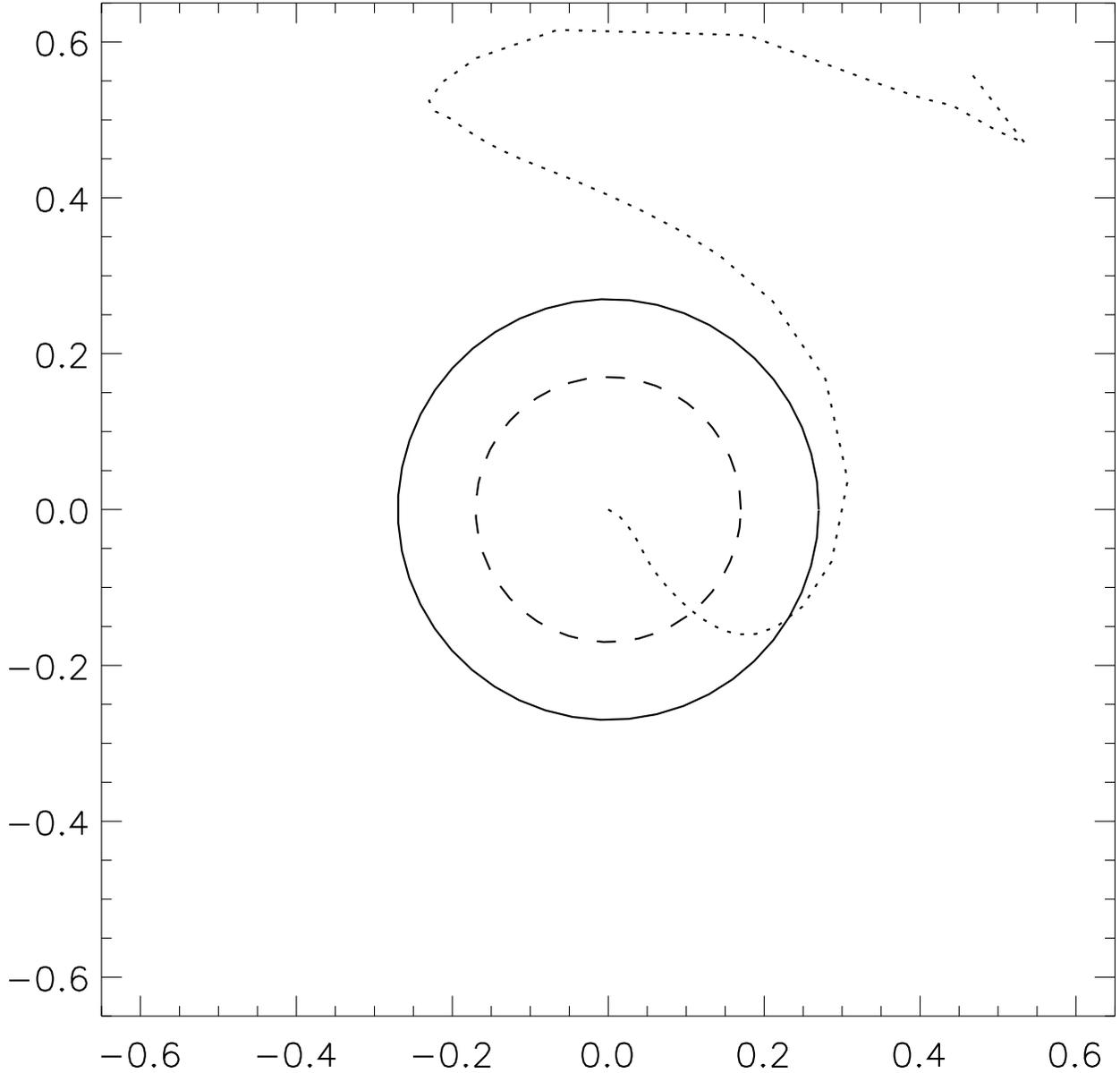} 
\caption{The dotted curve
displays the azimuthal structure $\phi_1(R)$ of the $m=1$
eigenfunction in the equatorial plane of model {\bf J250} at $t=95.4
T_c$ when GRR was turned on. The solid circle denotes the corotation
radius; the dashed circle denotes the location of density maximum.
\label{fig:phase_j250} }
\end{figure}

\begin{figure}[]
%%\centerline{
%%\includegraphics[width=3.5in]{phase.j068.ps}}
%%\epsscale{1.0} \plotone{phase.j068.ps}
\epsscale{1.0} \plotone{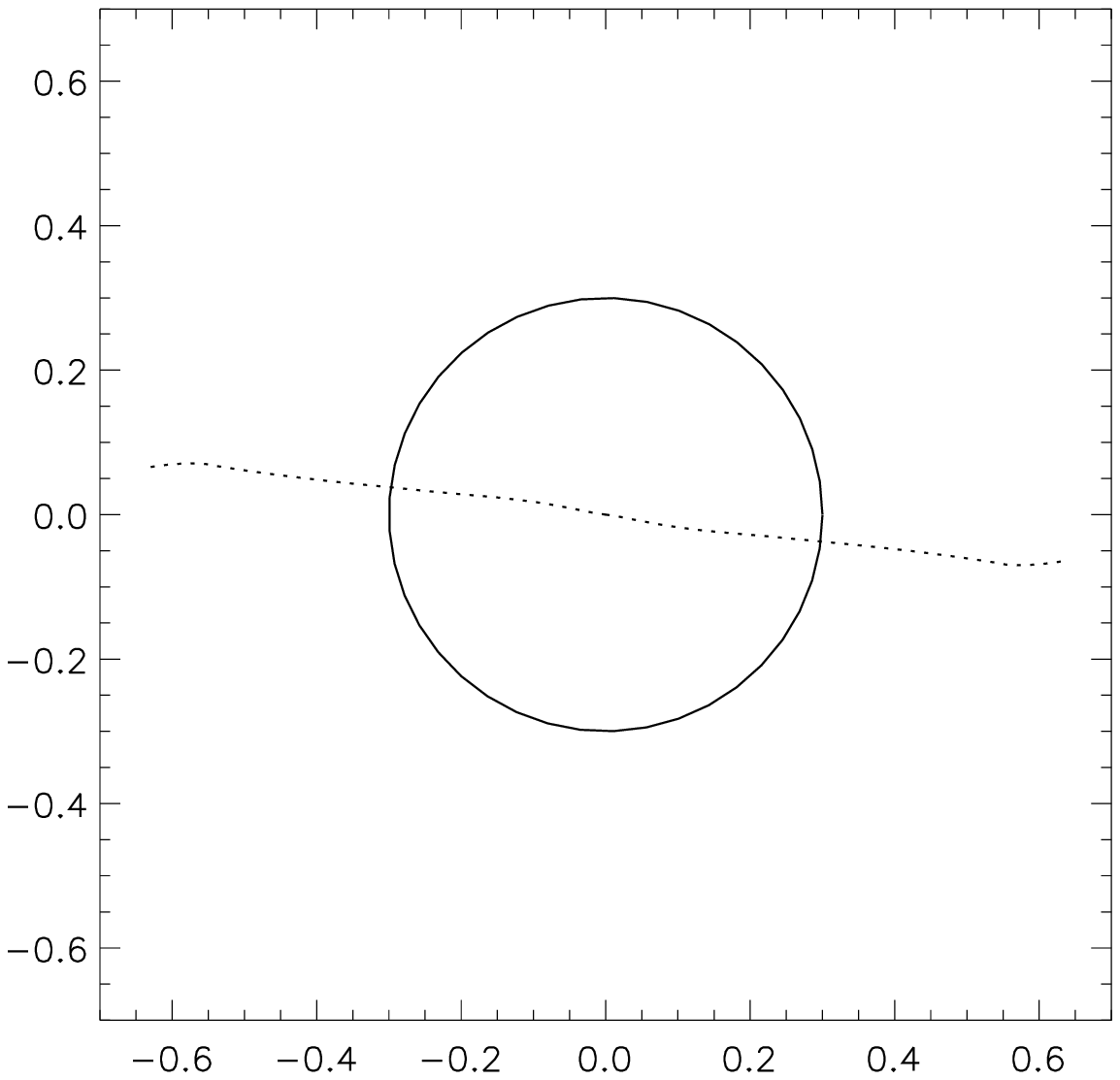}
\caption{Same as Figure \ref{fig:phase_j250} but for the $m=2$ mode
   of model {\bf J068} when $t=70 T_c$; there is no dashed circle
   because the density maximum lies at the center of the star.
  \label{fig:phase_j068} 
}
\end{figure}

Finally, in an effort to draw a further connection between our
results and previous analyses, we have examined the azimuthal
structure of the unstable eigenmodes in our models.  Figure
\ref{fig:phase_j250} displays the phase angle of the $m=1$ mode as a
function of $R$ in the equatorial plane of model {\bf J250}. The
pattern shows a slightly trailing segment in the innermost region,
then transforms itself into a leading wave outside the density
maximum, and finally a loosely wound, trailing segment appears again
outside the corotation radius. This particular characteristic is
reminiscent of the $m=1$ structure that was observed in an earlier
study of Papaloizou-Pringle modes by \cite{WTH94}, and it is quite
similar to the azimuthal structure of the one-armed spiral mode
discussed by \cite{saijo:05}. In contrast, Figure
\ref{fig:phase_j068} shows the phase-angle plot of the $m=2$ mode
for model {\bf J068}. No spiral structure is observed for the $m=2$
mode; instead, the phase angles at different radii are roughly
aligned along a straight line indicating a purely bar-like
deformation.
The radial eigenfunction of the $m=1$ mode shown in Fig. \ref{fig:drhoj250}
is not monotonic and has a node between the center and the surface;
we suspect it belongs to the family of Papaloizou-Pringle modes
that arises in pressure-supported tori and thick accretion disks.
The nodeless feature of the radial eigenfunctions
for $m=2, 3$ modes suggests that they might be f-mode oscillations
(this interpretation is consistent with the unstable mode structure
presented by \cite{SKE03});
but their structures are different in details:
the eigenfunction of the $m=2$ mode (seen in {\bf J068}) starts 
from zero at the center, monotonically increases outward,
and reaches a maximum amplitude near the surface;
in contrast, the eigenfunction of the $m=3$ mode (seen in {\bf J127})
has a maximum between the center and the surface.

%%One important issue related to the credibility of our results is
%%whether the movement of the center of mass of the system will give
%%rise to some spurious $m=1$ mode.  We have closely monitored the
%%motion of the center of mass of the system in all our simulations
%%and it appears to be well controlled near the origin (for example,
%%it is always confined to within the innermost radial grid zone).
%%Hence, it is unlikely that the one-armed $m=1$ instability observed
%%in our simulations is caused by the motion of center of mass. On the
%%other hand, if it is, it should give a systematic spurious $m=1$
%5mode in all the simulations, no matter whether the model is stable
%%or unstable, whereas, we didn't observe the $m=1$ mode in our stable
%%model.

\acknowledgments

We would like to thank Hui Li for insightful and useful
discussions at the 47th Meeting of the Division of Plasma Physics
of the American Physical Society. We also thank Juhan Frank,
Patrick Motl, Shin Yoshida, and Christian Ott for helpful comments and
suggestions. It is a pleasure to thank Anna Watts for useful
conversations at the Eighth Divisional Meeting of High Energy
Astrophysics Division of the American Astronomical Society. 
We thank an anonymous referee for helpful suggestions.
This work was partially supported by NSF grants AST-0407070 and
PHY-0326311.
The computations were performed on the Supermike cluster and
the Superhelix cluster at LSU, and on the Tungsten cluster
at National Center for Supercomputing Applications (NCSA).

\clearpage

%% The following command ends your manuscript. LaTeX will ignore any text


\begin{thebibliography}{}

\bibitem[Akiyama \& Wheeler(2005)]{AW05}Akiyama, S., \& Wheeler, J. C., 2005, \apj, 629, 414
\bibitem[Andersson(2003)]{A03} Andersson, N. 2003, Class. Quant. Grav., 20, 105
%%\bibitem[Arras et al.(2003)]{AFMSTW03} Arras, P., et al. 2003, \apj, 591, 1129
\bibitem[Brown(2000)]{B00}Brown, J. D. 2000, \prd, 62, 084024
\bibitem[Cazes \& Tohline(2000)]{CT00}Cazes, J. E., \& Tohline, J. E. 2000, \apj, 532, 1051
\bibitem[Centrella et al.(2001)]{CNLB01}Centrella, J. M., New, K.C.B., Lowe, L., Brown, J.D.
2001, \apj, 550, L193
\bibitem[Chandrasekhar(1969)]{Ch69}Chandrasekhar, S. 1969,
Equilibrium Figures of Equilibrium,  New Haven, CT:  Yale Univ.
Press
\bibitem[Chandrasekhar(1970)]{Ch70} Chandrasekhar, S. 1970, \apj,
161, 561
\bibitem[Drazin \& Reid(1981)]{DR81} Drazin, P. G., \& Reid, W. H., Hydrodynamic Stability
  (Cambridge: Cambridge Univ. Press)
\bibitem[Durisen et al.(1986)]{DGTB86} Durisen, R. H., Gingold, R. A., Tohline, J. E., \& Boss, A. P.
1986, \apj, 305, 281

\bibitem[Frank \& Robertson(1988)]{FR88}Frank, J., \& Robertson, J.A., 1988, \mnras, 232, 1

\bibitem[Friedman \& Schutz(1978)]{FS78} Friedman, J., \& Schutz,
B. F. 1978, \apj, 222, 281

\bibitem[Goldreich, Goodman, \& Narayan(1986)]{GGN86}Goldreich, P., Goodman, J., \& Narayan, R.,
1986, \mnras, 221, 339

\bibitem[Hachisu(1986)]{H86} Hachisu, I. 1986, \apjs, 61, 479
%%\bibitem[Hawley, Balbus \& Winters(1999)]{HBW99} Hawley, J. F.,
%%Balbus, S. A., \& Winters, W. F. 1999, \apj, 518, 394
%%\bibitem[Imamura, Friedman, \& Durisen(1985)]{IFD85} Imamura, J.
%%N., Friedman, J. L., \& Durisen, R. H. 1985, \apj, 294, 474
\bibitem[Ipser \& Lindblom(1990)]{IL90} Ipser, J. R., \&
Lindblom, L. 1990, \apj, 355, 226
\bibitem[Ipser \& Lindblom(1991)]{IL91} Ipser, J. R., \&
Lindblom, L. 1991, \apj, 373, 213

%%\bibitem[Kormandy \& Norman(1979)]{KN79} Kormandy, J., \&
%%Norman, C. A., 1979, \apj, 233, 539


\bibitem[Lai \& Shapiro(1995)]{LS95} Lai, D., \& Shapiro, S. L.
1995, \apj, 442, 259


\bibitem[Li et al.(2000)]{LFLC00}Li, H., Finn, J. M., Lovelace, R. V. E.,
\& Colgate, S. A., 2000, \apj, 533, 1023
\bibitem[Li et al.(2001)]{LCWL01}Li, H., Colgate, S. A., Wendroff, B., Liska, R., 2001, \apj, 551, 874


\bibitem[Liu(2002)]{L02} Liu, Y. T., 2002, \prd, 65, 124003

%%\bibitem[Lindblom \& Detweiler(1977)]{LD77} Lindblom, L., \&
%%Detweiler, S. 1977, \apj, 211, 565
%%\bibitem[Lindblom \& Detweiler(1979)]{LD79} Lindblom, L., \& Detweiler, S.
%%1979, \apj, 232, L101
%%\bibitem[Lindblom \& Hiscock(1983)]{LH83} Lindblom, L., \&
%%Hiscock, W. A. 1983, \apj, 267, 384
%%\bibitem[Lindblom \& Mendell(1995)]{LM95} Lindblom, L., \&
%%Mendell, G. 1995, \apj, 444, 804

\bibitem[Lovelace \& Hohlfeld(1978)]{LH78}Lovelace, R. V. E., Hohlfeld, R. G., 1978, \apj, 221, 51
\bibitem[Lovelace et al.(1999)]{LLCN99}Lovelace, R. V. E., Li, H., Colgate, S. A.,
Nelson, A. F., 1999, \apj, 513, 805

\bibitem[Motl, Tohline, \& Frank(2002)]{MTF02} Motl, P. M.,
Tohline, J. E., \& Frank, J. 2002, \apjs, 138, 121
\bibitem[Narayan, Goldreich, \& Goodman(1987)]{NGG87} Narayan, R., Goldreich, P., \& Goodman, J.,
1987, \mnras, 228, 1
\bibitem[New, Centrella \& Tohline(2000)]{NCT00}New, K. C. B., Centrella, J. M., \& Tohline, J. E.
2000, \prd, 62, 064019

\bibitem[Ott et al.(2004)]{ott:04}Ott, C. D., Burrows, A., Livne, E., \& Walder, R.
2004, \apj, 600, 834

\bibitem[{{Ott} {et~al.}(2005a){Ott}, {Ou}, {tohline}, \& {Burrows}}]{ott:05a}
{Ott}, C.D., {Ou}, S., {Tohline}, J.E., {Burrows}, A., 2005a, \apj, 625, L119

\bibitem[Ott et al.(2005b)]{ott:05b}Ott, C. D., Burrows, A., Thompson, T. A., Livne, E., \& Walder, R.
2005b, astro-ph/0508462


\bibitem[Ou, Tohline \& Lindblom(2004)]{OTL04}Ou, S., ,Tohline, J.E, Lindblom, L. 2004,
\apj, 617, 490


\bibitem[Papaloizou \& Pringle(1984)]{PP84}Papaloizou, J.C.B., Pringle, J.E., 1984,
\mnras, 208, 721
\bibitem[Papaloizou \& Lin(1989)]{PL89}Papaloizou, J.C.B., Lin, D. N. C., 1989,
\apj, 344, 645


\bibitem[{{Saijo} {et~al.}(2003){Saijo}, {Baumgarte}, \& {Shapiro}}]{saijo:03}
{Saijo}, M., {Baumgarte}, T.~W., \& {Shapiro}, S.~L. 2003, \apj, 595, 352
%%\bibitem[{{Saijo} {et~al.}(2005){Saijo}, \& {Yoshida}}]{saijo:05}
\bibitem[{{Saijo \& Yoshida}(2005){Saijo}, \& {Yoshida}}]{saijo:05}
{Saijo}, M., \& {Yoshida}, S., 2005, astro-ph/0505543

\bibitem[Shibata \& Karino(2004)]{SK04}Shibata, M., \& Karino, S. 2004, \prd, in press
(astro-ph/0408016)
\bibitem[Shibata et al.(2002)]{SKE02}Shibata, M., Karino, S., \& Eriguchi, Y. 2002, \mnras, 334,L27
\bibitem[Shibata et al.(2003)]{SKE03}Shibata, M., Karino, S., \& Eriguchi, Y. 2003, \mnras, 343, 619
%%\bibitem[Stergioulas(2003)]{S03}Stergioulas, N. 2003, Living Rev. Relativity, 6, 3
%%\bibitem[Stergioulas \& Friedman(1998)]{SF98} Stergioulas, N., \&
%%Friedman, J. L. 1998, \apj, 492, 301
\bibitem[Tassoul(1978)]{T78}Tassoul, J.-L. 1978, Theory of Rotating Stars,
Princeton: Princeton University Press
\bibitem[Tohline, Durisen, \& McCollough(1985)]{TDM85}Tohline, J. E., Durisen, R. H., \& McCollough, M.
1985, \apj, 298, 220
\bibitem[Tohline \& Hachisu(1990)]{TH90}Tohline, J. E., Hachisu, I., 1990, \apj, 361, 394

\bibitem[Watts et al.(2004)]{watts:04}Watts, A.~L., Andersson,
N., \& Jones, D.~I.\ 2005, \apjl, 618, L37

\bibitem[Williams \& Tohline(1988)]{WT88}
Williams, H. A., \& Tohline, J. E. 1988, \apj, 334, 449

\bibitem[Woodward et. al.(1994)]{WTH94}
Woodward, J. W., Tohline, J. E., \& Hachisu, I., 1988  \apj, 420,
247

%%\bibitem[Yoshida \& Eriguchi(1995)]{YE95} Yoshida, S., \& Eriguchi, Y. 1995, \apj, 438, 830
\end{thebibliography}
\end{document}